\documentclass[%
 reprint,
longbibliography,
 amsmath,amssymb,
 aps,
]{revtex4-1}

\pdfoutput=1

\usepackage{graphicx}
\usepackage{dcolumn}
\usepackage{bm}

\usepackage[usenames]{color}
\usepackage{url}
\usepackage{hyperref}

\begin{document}

\preprint{APS/123-QED}

\title{Unsupervised interpretable learning of topological indices invariant under permutations of atomic bands}

\author{Oleksandr Balabanov}
 
\author{Mats Granath}%

\affiliation{Department of Physics, University of Gothenburg, SE 412 96 Gothenburg, Sweden}%

\date{\today}

\begin{abstract}

Multi-band insulating Bloch Hamiltonians with internal or spatial symmetries, such as particle-hole or inversion, may have topologically disconnected sectors of trivial atomic-limit (momentum-independent) Hamiltonians. We present a {neural-network-based} protocol for finding topologically relevant indices that are invariant under transformations between such trivial atomic-limit Hamiltonians, thus corresponding to the standard classification of band insulators. The work extends the method of `topological data augmentation’ for unsupervised learning introduced in Ref.~\cite{balabanov1} by also generalizing and simplifying the data generation scheme and by introducing a special `mod’ layer of the neural network appropriate for $Z_n$ classification. Ensembles of training data are generated by deforming seed objects in a way that preserves a discrete representation of continuity. In order to focus the learning on the topologically relevant indices, prior to the deformation procedure we stack the seed  Bloch Hamiltonians with a complete set of symmetry-respecting trivial atomic bands. The obtained datasets are then used for training an interpretable neural network specially designed to capture the topological properties by learning physically relevant momentum space quantities, even in crystalline symmetry classes.

\end{abstract}

\maketitle

\section{Introduction}

To find and classify exotic phases of quantum matter is of key importance in modern science. These phases often point to materials with unique properties and high scientific impact. A particular type of quantum matter, the so-called topologically non-trivial quantum phases, are especially at the frontier of current research \cite{RevModPhys.82.3045,RevModPhys.83.1057, RevModPhys.88.035005}. Topological phase transitions lie outside the standard Landau paradigm of symmetry breaking and highlight topological non-triviality of the underlying state spaces. Despite of all the progress made and understanding gained over the years, the field is still very far from being complete.

Although still in its infancy as applied to topological matter, the use of Machine Learning (ML), specialized in big data pattern recognition, has the potential to revolutionize the field by circumventing the need for explicit analytic representations of topological markers (indices). Ideally, identifying topology using ML can subsequently guide further theoretical developments.   
In many areas of quantum physics, applications of ML has become a major research area \cite{dunjko2018machine,carrasquilla2020machine}, for example to represent many-body states \cite{carleo2017solving,Gao2017, PhysRevX.7.021021, PhysRevB.96.195145, PhysRevB.96.205152, PhysRevB.97.195136, PhysRevB.99.165123, melko2019restricted, PhysRevX.8.011006, PhysRevLett.121.167204}, identifying both symmetry breaking and topological phase transitions \cite{PhysRevLett.120.066401, PhysRevB.98.085402, PhysRevE.99.023304, PhysRevB.97.115453, Ming2019, Greplova_2020, 2019arXiv190904784T, 2019arXiv190103346C, PhysRevResearch.2.023283, Carrasquilla2017, PhysRevX.7.031038, Broecker2017, PhysRevB.97.045207, PhysRevB.97.134109, PhysRevE.99.032142, PhysRevB.96.245119, Rem2019, PhysRevB.99.104106, 2019arXiv190111042G, van2017learning,rodriguez2019identifying, PhysRevLett.124.226401,PhysRevLett.124.185501,che2020topological, PhysRevB.101.064406},
for quantum state tomography~\cite{Torlai2018, PhysRevLett.118.216401, PhysRevLett.123.230504}, quantum control \cite{PhysRevX.8.031079,PhysRevX.8.031086,dalgaard2020global} and quantum error correction \cite{PhysRevLett.119.030501,Baireuther2018machinelearning, 2018arXiv181007207S, Andreasson2019quantumerror,nautrup2019optimizing,PhysRevResearch.1.033092}. 

\begin{figure} \centering
    \includegraphics[width=8.5cm,angle=0]{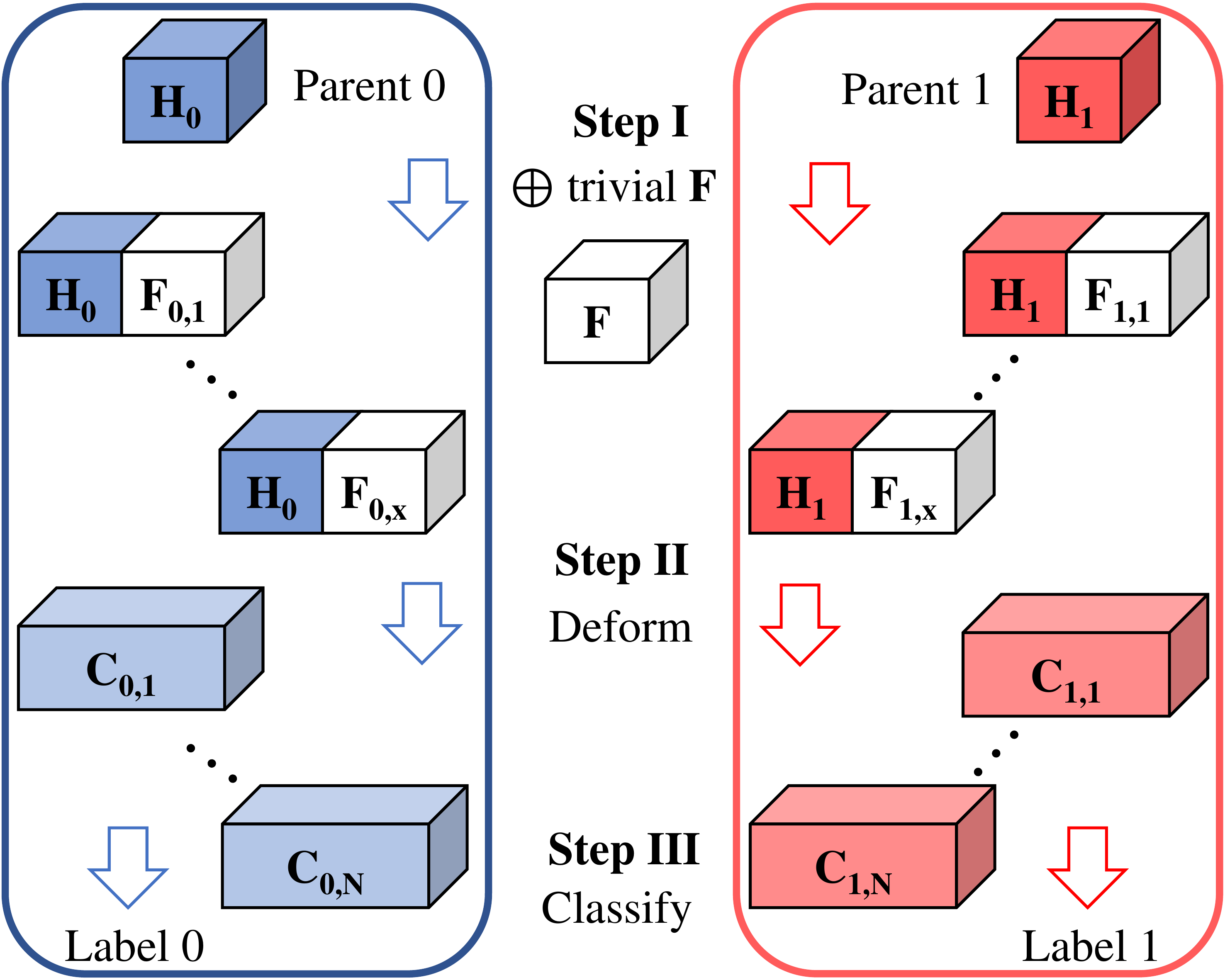}
    \caption{Protocol for generating the training data: {Step~I}. The parent Hamiltonians $H_i(k)$ with $i = 0, 1$ are extended to $H_i(k) \oplus F_{i, j}(k)$, where $F_{i, j}(k)$ are randomly generated trivial atomic-limit Bloch Hamiltonians.  {Step~II}. The {stacked} parents are augmented via continuous deformations.  {Step~III}. The obtained datasets, labeled `0’ and `1’, are used for training a neural network specifically designed for capturing topological indices. }
     \label{fig:fig1}
\end{figure}

The use of Artificial Neural Networks (NNs) or Deep Learning, in particular, stand out as the premier ML tool, with the power both to find complex patterns in data and to generalize this knowledge to previously unseen data. The capabilities and limitations of NNs applied to the task of classifying phases of topological quantum matter are currently under active research~\cite{PhysRevLett.120.066401, PhysRevB.98.085402, PhysRevE.99.023304, PhysRevB.97.115453, Ming2019, Greplova_2020, 2019arXiv190904784T, 2019arXiv190103346C, PhysRevResearch.2.023283}. A potential limitation to using NNs is that common protocols are based on supervised learning which requires 
labeled  data (object and corresponding topological marker) from an explicit model. This limits the prospectives for using ML to explore unknown topological structures. To begin to address this issue several \textit{unsupervised} protocols for doing topological classifications  have recently been developed  \cite{van2017learning,rodriguez2019identifying,balabanov1, PhysRevLett.124.226401,PhysRevLett.124.185501,che2020topological}. The method presented by us in Ref.~\cite{balabanov1} is based on `topological data augmentation’, where datasets of topologically equivalent children states are created from single parent states. The created data is then used for training a NN, using the unsupervised `learning by confusion’ concept introduced in Ref.~\cite{van2017learning}: Only datasets that have some distinguishing feature can be functionally classified by the network. Thus, any two a priori unknown datasets can be used for training the network with dummy label targets, with successful training outcome only if the two sets are topologically distinct. The procedure also used interpretable neural networks, designed to learn an intermediate momentum space output which is closely related to integral expressions over the local curvature.  
A crucial and challenging requirement for the procedure to work is that the data is sufficiently randomized to erase any non-topological information related to the specific parent states. In order to accomplish this randomization without destroying the topological information we construct in Ref.~\cite{balabanov1} explicit topology preserving transformations, valid in the discretized Brillouin zone. However, to construct such transformations, one might argue, requires a good understanding of the topological structure. Thus to relax the specificity of the topology preserving deformations would be a step forward in the spirit of the outlined objective.

With this in mind, in this paper we significantly advance the protocol from Ref.~\cite{balabanov1}  and reveal new horizons of its applicability. As before, the protocol is split into two main stages, data generation and training of the network. The training data is generated from two parent states via random topology-preserving deformations. Here we always require the states to change slowly with momentum, in this way automatically ensuring continuity of the applied deformations. The procedure to do this while at the same time ensuring sufficient exploration of the space in order to obscure any non-topological features is one of the main methodological points. This also allows us to consider more general quantum systems without requiring any external knowledge on the topological structure of the underlying state space. We also modified the network's layout allowing the net to generically represent a much broader class of topological indices, including such that are sensitive to high symmetry points, lines, and surfaces in the Brillouin zone.

An additional limitation of the original protocol of topological data augmentation is that network may learn momentum space local invariants that are irrelevant for establishing non-trivial topological features. As an example, consider Hamiltonians with inversion symmetry, where in the atomic limit the individual orbitals may be odd or even under inversion. A half-filled 4-band Hamiltonian in the atomic limit may have two occupied bands either both with even parity, both with odd parity, or one band with even and one with odd. These three sectors of atomic Hamiltonians are topologically trivial, have no edge states~\cite{Hughes_Inversion}, but nevertheless cannot be connected by continuous non gap-closing deformations. Thus, a network trained on dataset based on deformations of  single seed objects from one of these three sectors may learn to distinguish different trivial band insulators based on a local index (such as the parity at the Brillouin zone center) rather than trivial from non-trivial, depending on the specifics of the two training datasets.

In order to address this issue we extend the Hilbert space and corresponding matrix dimensions of the Bloch Hamiltonians that we want to classify from $n$ to $2n$ bands,  by {stacking them with} $n$ atomic bands as sketched in Fig.~\ref{fig:fig1}: The children states are produced by first {stacking} the parents {with} a number of trivial atomic bands and only then deforming them employing random topology-preserving deformations. The created data is then used for training an interpretable NN allowing us to extract the learned topological quantities. Trivial atomic bands here mean bands that respect the symmetries of the corresponding symmetry class but do not vary in momentum, they represent the intrinsically trivial atomic limit where individual atomic bands do not hybridize. Thus, within the new procedure we purposely assign the same labels to  any Bloch Hamiltonians differing by trivial atomic bands, in this way penalizing the NN for outputting topologically irrelevant indices that  differentiate between \textit{a priori} trivial objects. This approach is inspired by the commonly employed \text{$K$-theoretic} analytic classification schemes \cite{Freed2013, doi:10.1063/1.3149495, RevModPhys.88.035005}, where the topological equivalence between two band insulators is defined up to stacking of trivial bands. All trivial Hamiltonians here constitute a trivial {monoid} element and any Hamiltonian can then be {stacked with} atomic bands without changing the topological class. 


The paper is organized as follows: The protocol is described in details in Sec.~II, including how to generate parent states that vary sufficiently slowly on the scale of discretized momentum, how these are {stacked} with atomic bands and subsequently deformed in a manner that conserves the discrete measure of continuity of the Bloch Hamiltonians. The protocol is exemplified on one-dimensional (1d) 4-band insulators from three different symmetry classes in Sec~III. This includes examples with particle-hole symmetry and inversion symmetry where the network learns to single out the high symmetry points in the Brillouin zone and combine these in the appropriate way to produce the topological index. A summary and outlook is given in Sec. IV.

\section{Methods~\label{sec:Methods}}

To set the stage, we describe our protocol in general terms and provide the details applicable for all examples to follow. Any one-dimensional (1d) gaped $n$-band system can be represented using a $n$ by $n$ Hermitian Bloch Hamiltonian $H({k})$ that is a periodic continuous function of momentum ${k}$. Two Bloch Hamiltonians are said to be \textit{topologically equivalent} if the occupied spaces can be continuously deformed into each other. Any Bloch Hamiltonian can therefore be continuously transformed to have energies $-1$ and $1$ for occupied and empty bands respectively, and we consider the space of normalized Bloch Hamiltonians only. The aim is then to find a topological index, call it $\nu$, labeling the Bloch Hamiltonians $H({k})$ by their topological equivalence classes. It is known that the space of all Bloch Hamiltonians in 1d is topologically trivial but the classification becomes nontrivial after imposing various symmetries on the systems, in this way restricting the allowed deformations. In this paper, we always assume the systems to be half filled and explicitly normalize the Bloch Hamiltonians to have equal number of positive and negative bands.

\begin{figure} \centering
    \includegraphics[width=8.5cm,angle=0]{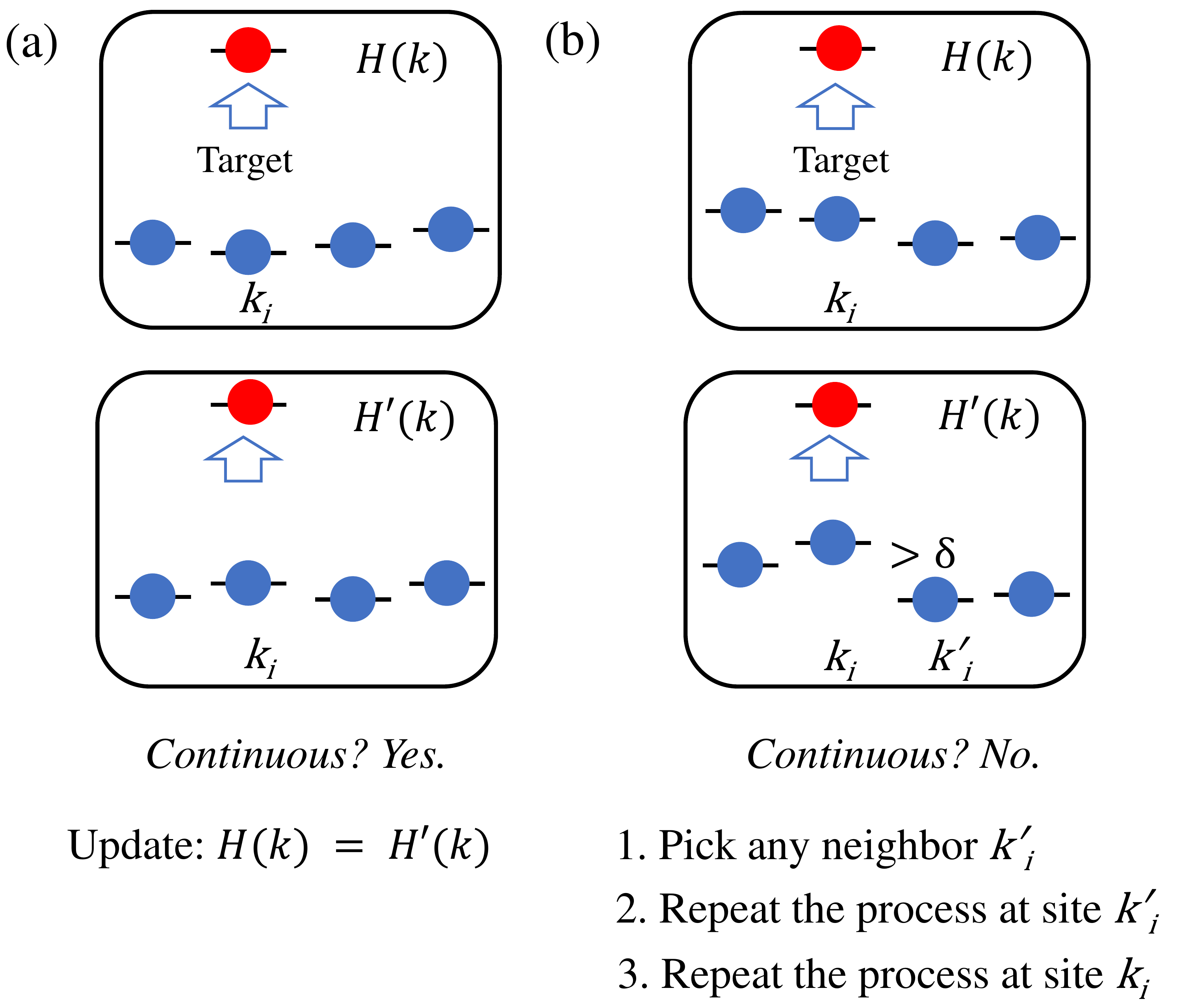} 
    \caption{One recursive iteration: The goal is to move the given Hamiltonian $H(k)$ at momentum site $k_i$ towards the target. First, we compute $H^\prime(k)$ obtained from $H(k)$  by doing one gradient descent step at site $k_i$. There are now two \text{scenarios}: (a)~$H^\prime(k)$ is continuous with respect to the continuity measure~$\delta$. Then, we update $H(k) = H^\prime(k)$. (b) $H^\prime(k)$ is discontinuous. Then, we choose any discontinuity neighboring point, say $k_i^\prime$, and do one recursive iteration at $k_i^\prime$. Next, one recursive iteration at $k_i$ is performed again.}
     \label{fig:fig2}
\end{figure}
\subsection{Prerequisites~\label{sec:Prerequisites}}

For our protocol to be functional we need to choose a distance measure between any Hermitian matrices~$H_1$ and~$H_2$ to later establish a notion of continuity for discretized Bloch Hamiltonians. The distance is here defined through relation
\begin{equation}
    d(H_1, H_2) = 1 - \frac{1}{N_{\text{o.b.}}}\sum_{i,j \, \in \, \text{o.b.}} |\langle \psi_{1,i} | \psi_{2,j} \rangle)|^2,
    \label{eq:distance}
\end{equation}
where $| \psi_{1/2,i} \rangle$ are the normalized eigenstates of $H_{1/2}$, with $i$ running over the $N_{\text{o.b.}}$ occupied bands of $H_{1/2}(k)$. Note that the distance $d(H_1({k}), H_2({k}))=0 $ for every momentum ${k}$ if and only if the (normalized) Bloch Hamiltonians $H_1({k})$ and $H_2({k})$ are identical. 

We also need an efficient algorithm to numerically move between the matrices and for this purpose we describe a procedure for gradually changing matrix $H_1$ towards $H_2$. The idea is to numerically minimize the distance $d(H_1, H_2)$ employing the gradient descent algorithm~\cite{PhysRevLett.124.226401}. Explicitly, in each numeric step we perform a unitary rotation $H_1 \rightarrow U_j H_1 U^\dagger_j$ using different unitary generators $U_j = \exp (i \phi_j \Lambda_{j})$, where $\Lambda_{j}$ is a basis of Hermitian matrices. The phase constants $\phi_j$ are computed using the gradient descent rule
\begin{equation}
    \phi_j = - \eta \, \frac{\partial \, d (U_j H_1 U^\dagger_j , H_2)}{\partial \phi_j}\Big |_{\phi_j=0}
    \label{eq:GD}
\end{equation}
with some small learning rate $\eta$. To avoid rapid changes possibly leading to discontinuous deformations we also restrict $\phi_j$ not to exceed some upper bound value $\phi_\text{max}$. The symmetries are maintained by doing the unitary rotations in pairs bonded by the symmetries, in this way forbidding the deformed Hamiltonian to go outside the considered symmetry class.

\begin{figure} \centering
    \includegraphics[width=8.5cm,angle=0]{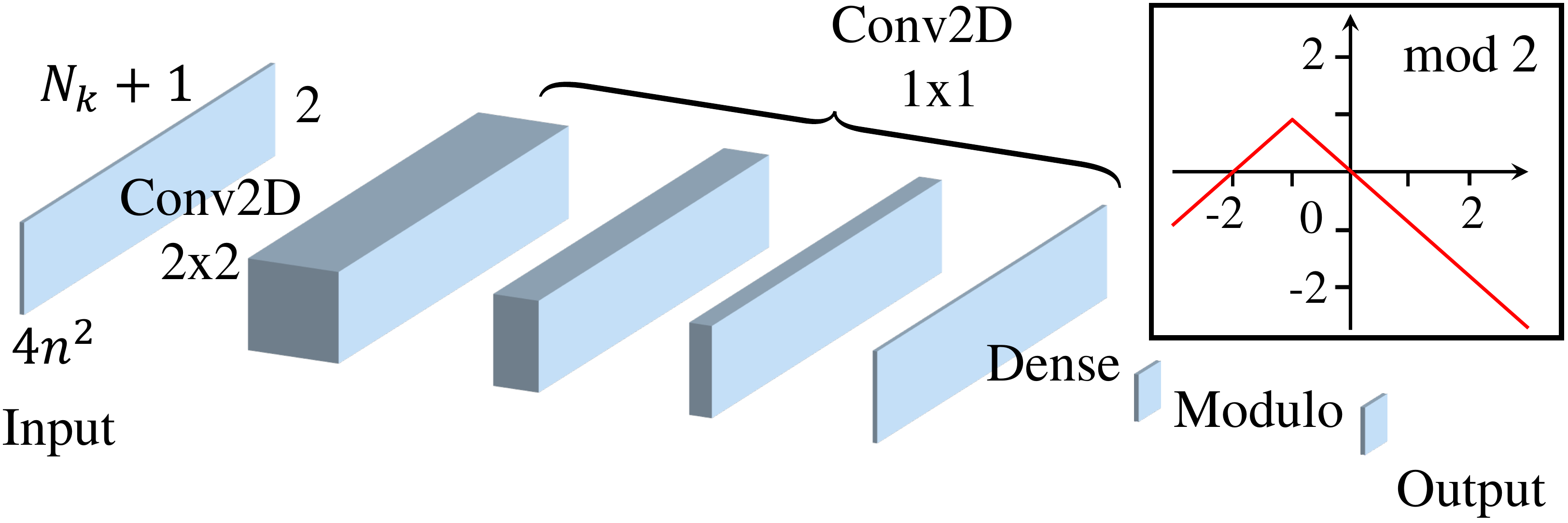}
    \caption{A generic class of neural networks designed  for interpretably classifying 1d band insulators. The input is fed in format $2 \times (N_k + 1) \times 4\, n^2$, where $N_k$ and $n$ are the numbers of momentum points and internal bands, respectively. The net consists of a Conv2D layer with receptive field of size (2,2) (over neighboring k-points and real and imaginary part of matrix entry, with 64 channel depth corresponding to each matrix entry), several Conv2D layers with receptive field of size (1,1), a Dense layer, and an optional Modulo layer. The last convolution layer outputs a single receptive field, corresponding to some momentum resolved local measure, which is key to the interpretability, as exemplified in Fig 4b, 5b, and 7.   The Modulo layer applies a predefined continuous function coinciding with the modulo operation in some key integer points: An example of such function mimicking mod 2 operation is depicted as well.}
    \label{fig:1D_network}
\end{figure}

\subsection{Generating parent states~\label{sec:Gener_Data}}

The training data is generated from two $n$ by $n$ parent states $H_0(k)$ and $H_1(k)$ defined on a grid of $N_k$ momentum points. The baseline trivial state $H_0(k)$ is taken to be any trivial atomic-limit Bloch Hamiltonian satisfying the symmetries. It is produced by repeating a random properly symmetrized $n$ by $n$ Hermitian matrix. The parent $H_1(k)$ is generated by symmetrizing a Bloch Hamiltonian 
\begin{equation}
    H(k) = \sum^{m}_{p=0} A_n \sin(p k) +  B_n \cos(p k)
    \label{eq:state_1}
\end{equation}
with $A_p$ and $B_p$ random $n$ by $n$ Hermitian matrices. With this procedure we can efficiently sample the symmetry classes and control state's continuity by choosing $m \ll N_k$. The continuity here is realized by requiring all neighbor to neighbor distances $d(H(k),H(k+\Delta k))$ (with $\Delta k=2\pi/N_k$) to be below some small threshold~$\delta$. Thus, the parent $H_1(k)$ is suggested to be randomly generated via Eq.~(\ref{eq:state_1}). Note, however, that in general we are mainly interested in picking $H_1(k)$ that is topologically nontrivial and therefore can facilitate the protocol by applying all available external intuition for purposely choosing a promising candidate for $H_1(k)$. Completely random guessing, however, is also anticipated to eventually lead to the same results according to learning by confusion ideology but one will need to make several attempts until reaching success in training. 

\subsection{Stacking parents with trivial atomic bands~\label{sec:Gener_Data1}}

Within our protocol of topological classification, Fig.~\ref{fig:fig1}, the parents $H_{i}(k)$ with $i = 0,1$ are first extended to $H_i(k) \oplus F_{i, j}(k)$ using random trivial atomic-limit Bloch Hamiltonians $F_{i, j}(k)$, and only then the {stacked} parents are deformed. The extension of the parents is not generally necessary within our protocol, however, the datasets generated in this way facilitate the NN to only look at {topologically relevant} indices that do not distinguish between different trivial atomic-limit Hamiltonians. We therefore develop a procedure for randomly generating trivial atomic-limit Bloch Hamiltonians $F(k)$ respecting the relevant symmetries. Each trivial Bloch Hamiltonian $F(k)$ is constructed by repeating some base $n$ by $n$ matrix $\tilde{F}$. The matrix $\tilde{F}$ has to respect the relevant symmetries, and within our protocol it is generated by filling the matrix entries with random complex numbers $(a + ib)$ with $a, b\in[-1, 1]$ (with real numbers $c\in[-1, 1]$ for diagonal entries), symmetrizing the obtained matrix to respect the given symmetries, and then normalizing it to have eigenvalues $\pm 1$. Depending on symmetry class the space of all symmetry-respecting matrices can be composed of multiple topologically disconnected sectors and these sectors can in general be non-uniformly represented in our randomly created data. Note that trivial Hamiltonians $F(k)$ corresponding to $\tilde{F}$ from disconnected sectors generically cannot be connected using continuous symmetry-preserving deformations. This may mislead our NN to find {topologically relevant} quantities because some connected parts of trivial atomic-limit Hamiltonians might be underrepresented. For avoiding our algorithm giving preference to a particular disconnected matrix sector we put the generated matrices into datasets, each composed of $10^3$ continuously connected random matrices. The connectivity is established by checking if they can be connected via the gradient descent algorithm described in Sec.~\ref{sec:Prerequisites} without breaking the symmetries, and the number of obtained disconnected sectors of trivial Hamiltonians depends on the symmetry class. We then pick a base matrix $\bar{F}_{i, j}$ defining $F_{i, j}(k)$ with equal probability from the created datasets of matrices $\tilde{F}$. In this way all the disconnected sectors are represented equally within the process. Using this approach we produce $j=1,...,N_{p}$ {{\em stacked}} parent states $H_i(k) \oplus F_{i, j}(k)$ corresponding to two parents $H_0(k)$ and $H_1(k)$. By using these trivially {stacked} parents we hint the network to learn only quantities that cannot differentiate between two trivial atomic states even if they cannot be continuously connected.  

\subsection{Topological data augmentation~\label{sec:Gener_Data2}}
The {stacked} parent states are then augmented by performing random deformations while keeping the states continuous and maintaining the symmetries. The deformations are implemented by doing unitary rotations of discretized $2n$ by $2n$ Bloch Hamiltonians $H(k)$ at distinct momentum points $k_i$ towards randomly generated target matrices $\tilde{H}$. These rotations are done in small steps by minimizing the distance function $d(H(k_i), \tilde{H})$ following gradient descent algorithm with some learning rate~$\eta$ and rotation cutoff~$\phi_\text{max}$. We also perform the steps in symmetry-bonded pairs to always maintain the symmetries. The unitary matrices are taken to be $U_j = \exp(i \psi_j \Lambda_j)$ with $\Lambda_j$ a set of basis matrices for $2n$ by $2n$ Hermitian matrices. Importantly, the state is kept continuous by recursively moving also matrices at neighboring momentum sites towards the same target once the distance exceeds the threshold~$\delta$, Fig.~\ref{fig:fig2}. In this way any deformation at one momentum site pulls the neighbors along with it, resulting in efficient augmentation procedure that at all times maintains continuity of the state. We terminate the deformation algorithm if: the corresponding matrix reached the target, the state could not be maintained continuous after a number of recursive steps $M_\text{max}$, or if we exceeded a maximum number of gradient descent learning steps $N_\text{max}$. The motivation for this procedure, which is similar to a string being pulled over a large distance at one point, is to be able to traverse large distances in the topological sector in an unbiased fashion. This in contrast to making small random local deformations, just producing a random walk, which in a high dimensional space will provide less efficient exploration.

\subsection{Neural network structure and training~\label{sec:Train}}

The two datasets  corresponding to topologically augmented stacked parents are then used to train a neural-network-based classifier specifically designed to represent generic expressions of topological indices. The NN is trained on the Bloch Hamiltonians $H(k)$ transformed from  $N_k \times 2n \times 2n$ complex-valued format to $2 \times (N_k + 1) \times 4\, n^2$ float format with an extra momentum site added to encode the  periodicity. The classifying network consists of several convolution layers and in the earlier work \cite{balabanov1} followed by a summation layer. The activation functions associated with each convolution layer are rectified linear units (relU) except the last one with linear activation. This network type was designed to perform identical local operations on each site and then sum the obtained outputs over the whole sample, in this way capturing topological numbers that are given by an integral over some local curvature. However, in presence of certain symmetries, high symmetry momentum points, {lines or surfaces} may be of special importance and the indices are expected to take a different form. We therefore advance the design of the network to cover a much broader class of expressions by changing the last summation layer to a fully connected (dense) layer, see Fig.~\ref{fig:1D_network}. To be concrete, we use a dense layer with a single output node, absolute weight values~$\leq 1$, zero bias, and linear activation. The reason for avoiding non-linear activation and large weights is to constrain the preceding feature map (the output from last convolution layer) to learn  relevant and {\em interpretable} momentum space quantities. Note that this family of networks, the net in Fig.~\ref{fig:1D_network} and its generalizations to 2d and 3d, can adjust to calculate sums of some local functions over arbitrary symmetry-preserving points, lines, or planes, and then add or subtract them in any order. A coarse grained version of the output before summation in this single node is what is displayed as the momentum resolved images in Figs.~4, 5, and 7. The network thus processes the local information in k-space convoluted over all bands in the first convolution layer (2 by 2 filters operate on nearest neighbor sites in k-space and real and imaginary part), which after subsequent non-linear operations outputs the relevant momentum resolved local quantities.

To capture an even larger class of indices we also suggest to use an extra layer applying a predefined operation on the single-valued output to represent a modulo function over some assumed range of integers. The modulo function itself was found to be not applicable because it is periodic and discontinuous, giving convergence problems during the training. A way around this problem is to use a continuous function coinciding with the modulo operation in some key integer points. An example mimicking mod 2 operation is given inside a box in Fig.~\ref{fig:1D_network}: This function outputs 0 for input 0 and -2, and outputs 1 for input -1. The network illustrated in Fig.~\ref{fig:1D_network} can be generalized to 2d case by analogously modifying the 2d network of Ref.~\cite{balabanov1}.




\section{Results}

Here we present results of the analysis for three examples of 4-band Hamiltonians in 1d, having chiral symmmetry, inversion symmetry, and particle-hole symmetry respectively. We show how the network correctly learns to classify trivial from non-trivial datasets in a way that gives  interpretable information corresponding to relevant momentum resolved quantities.

\subsection{Hyperparameters~\label{sec:1D_0}}

We selected the same hyperparameters throughout all of our examples and they are listed here. The momentum space was discretized by $N_k = 100$ points. The gradient descent leaning rate was taken to be $\eta = 0.1$, the rotation cutoff~$\phi_\text{max} = 0.1$, and the continuity parameter $\delta = 0.1$. The maximum numbers of performed learning and recursive steps before terminating the algorithm were $N_\text{max} = 20$ and $M_\text{max} = 20$, respectively. The trivial parent $H_0(k)$ was generated by picking a random symmetry-respecting matrix and duplicating it for all $N_k$ points. The second parent $H_1(k)$ was randomly generated by symmetrizing the output of Eq.~(\ref{eq:state_1}) with $m = 4$. The two parents were then stacked with $N_{p} = 10$ randomly chosen symmetry-respecting {trivial atomic-limit} Hamiltonians to create two sets of $2n$ band Bloch Hamiltonians. Each of the {stacked} parents was then used to create $10^3$ children. For doing so, at first we deformed them by $10$ independent single-site unitary rotations, and only then saved a child for each new unitary deformation. In this way we produced two datasets of $10^4$ children Bloch Hamiltonians corresponding to two original parents. 


\subsection{Multiband 1d insulators with chiral symmetry~\label{sec:1D_1}}

\begin{figure} \centering
    \includegraphics[width=8.5cm,angle=0]{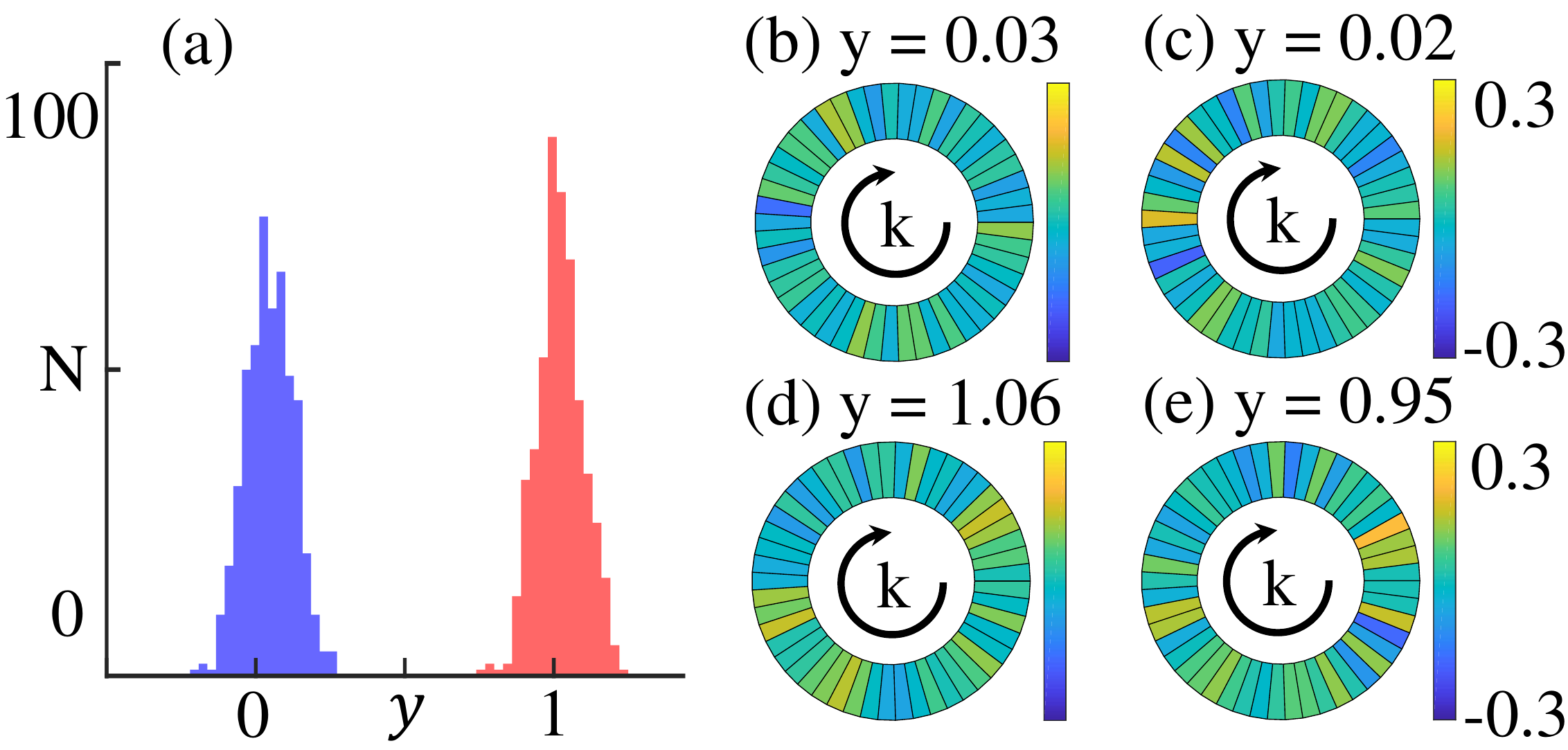}
    \caption{Topological classification of 1d Bloch Hamiltonians with chiral symmetry:  a) The net's output (y) evaluated on a test dataset of $10^{3}$ samples. The data corresponding to the trivial reference (randomly selected) parent is depicted in blue (red). b) - e) Local quantities and corresponding to them outputs exemplified on 4 randomly selected systems. }
    \label{fig:AIII_topo_numbers_test}
\end{figure}

As our first illustration, we implement the protocol for topological classification of 4-band insulators in 1d with chiral symmetry. It is well known that all gapped systems within this symmetry class are distinguished by a $Z$ topological invariant, the so-called winding number~\cite{RevModPhys.88.035005}. Here we train a NN to fit into a representation of the winding number without using any externally labeled data: The training data is entirely produced via the topological data augmentation procedure described in Sec.~\ref{sec:Gener_Data}. This symmetry class was also explored for 2-band systems in Ref.~\cite{balabanov1}, using a different protocol for generating the training data, but with similar results. 

Bloch Hamiltonians satisfying chiral symmetry must obey $H(k) = -U_S^\dagger \, H(k) \, U_S$, with a unitary matrix $U_{S}$ chosen to be a diagonal matrix with first half of entries $+1$ and last half of entries $-1$, thus consisting of off-diagonal blocks. In order to symmetrize randomly generated Hermitian matrices $A$ and Bloch Hamiltonians $H(k)$ we then performed the following operations on them
\begin{align}
    A &\rightarrow A - U_S \, A \,U_S^\dagger \\
    H(k)&\rightarrow H(k)- U_S\, H(k) \,U_S^\dagger. 
    \label{eq:chiral}
\end{align}
It was found that all generated symmetrized matrices $A$ could be connected via the gradient descent algorithm described in Sec.~\ref{sec:Prerequisites}, creating a single dataset of interconnected matrices, in agreement with our anticipation. The base matrices defining the chiral atomic-limit Bloch Hamiltonians were therefore uniformly picked from that single dataset, taken to be sampled by $10^3$ matrices. (Note that, because there is only a single trivial atomic sector, the stacking procedure is not strictly necessary for this class of systems.) All other specifications needed for topological data augmentation are listed in Sec.~\ref{sec:1D_0}.

At the training stage we employed a NN with a layout thoroughly described in Sec.~\ref{sec:Train} and depicted in Fig.~\ref{fig:1D_network}. The convolutional part of the network was explicitly taken to consist of one 2d convolution layer of 512 filters with (2, 2) receptive fields and three 2d convolution layers of 256, 128, 1 filters with (1, 1) receptive fields.  The network was successfully trained without any modulo layer in this case, confirming that the learned topological number is a $Z$ invariant. In total our network has $296 \, 038$ trainable parameters. 
\begin{figure} \centering
    \includegraphics[width=8.5cm,angle=0]{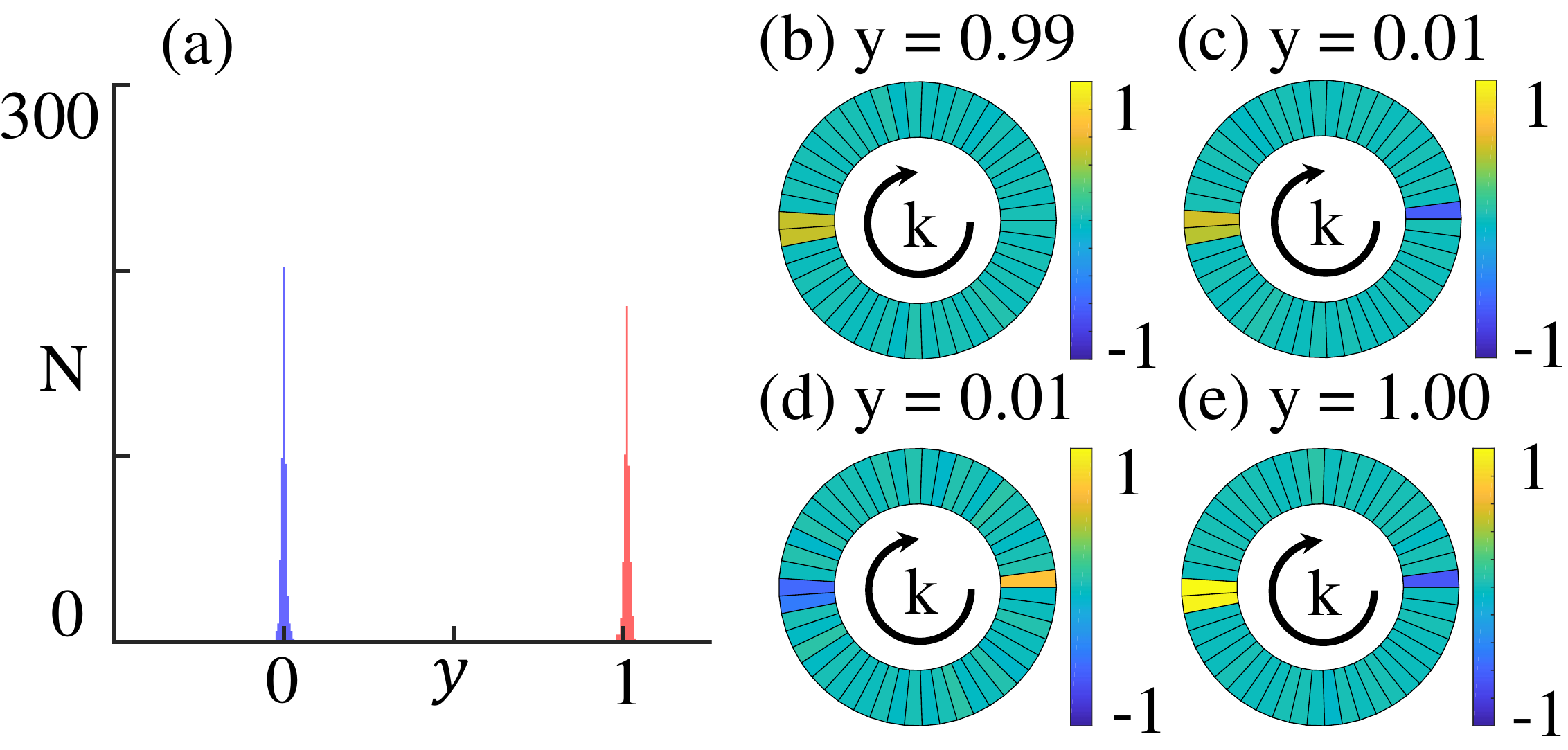}
    \caption{Topological classification of 1d Bloch Hamiltonians respecting inversion:  a) The output (y) of the network tested on a dataset of $10^{3}$ children systems. The data corresponding to the trivial reference and randomly selected parents is pictured in blue and red, respectively. b) - e) Local quantities and corresponding to them outputs associated with 4 distinct random systems. }
    \label{fig:IS_topo_numbers_test}
\end{figure}

The training was done on $2 \cdot 9500$ and tested on $2 \cdot 500$ samples using absolute mean error cost and Adam optimizer. We trained over $2000$ epochs with learning rate $\eta = 10^{-4}$. Importantly, we also effectively augmented the training dataset: Before each epoch every Bloch Hamiltonian was translated by a random number of momentum sites and uniformly rotated using a random symmetry-respecting unitary operator.

The results are presented in Fig.~\ref{fig:AIII_topo_numbers_test}. The network has successfully learned to distinguish the datasets corresponding to two ensembles of topologically equivalent children, Fig.~\ref{fig:AIII_topo_numbers_test}a: The classification accuracy evaluated on the test set is $100\%$. Importantly, the NN layout allows us to interpret the obtained classification and find a local quantity corresponding to the learned topological index. In Fig.~\ref{fig:AIII_topo_numbers_test}b - e we exemplified the learned local quantity on four representative systems from the test dataset.


\subsection{Multiband 1D insulators with inversion symmetry~\label{sec:1D_2}}

In our second example we focus on 4-band systems with inversion symmetry and implement our protocol for classifying them. The topological phases within this symmetry class are characterized by a $Z$ invariant $\nu_{IS} = n_{0} - n_\pi$ with $n_{0/\pi}$ counting a number of negative parity eigenstates at high-symmetry points $k = 0$ and $k = \pi$~\cite{2014arXiv1403.5558L, Hughes_Inversion}.

The inversion relation reads as $H(k) = U_{IS}^\dagger \, H(-k) \, U_{IS}$,  where $U_{IS}$ is some unitary matrix and it is here chosen to be the same as $U_S$, i.e. a diagonal matrix with $+1$ and $-1$ entries. The symmetrization of randomly generated Hermitian matrices $A$ and Bloch Hamiltonians $H(k)$
is done via
\begin{align}
    A &\rightarrow A + U_{IS} \, A \, U_{IS}^\dagger \\
    H(k)&\rightarrow H(k) + U_{IS} \, H(-k)\,  U_{IS}^\dagger. 
    \label{eq:IS}
\end{align}
The generated symmetrized matrices $A$ were found to create three distinct blocks of matrices, with all matrices connected via the gradient descent algorithm within each block. Analytically, as discussed in the introduction, the disconnected blocks correspond to three different combinations of inversion-symmetric eigenvalues $\pm 1$ of the  occupied eigenstates. Thus, for producing inversion-symmetric trivial atomic-limit Hamiltonians we uniformly picked matrices from these three blocks. Each block was taken to consist of $10^3$ generated matrices. Other details of data generation are given in Sec.~\ref{sec:1D_0}.

\begin{figure}[t!]  \centering
    \includegraphics[width=8.5cm,angle=0]{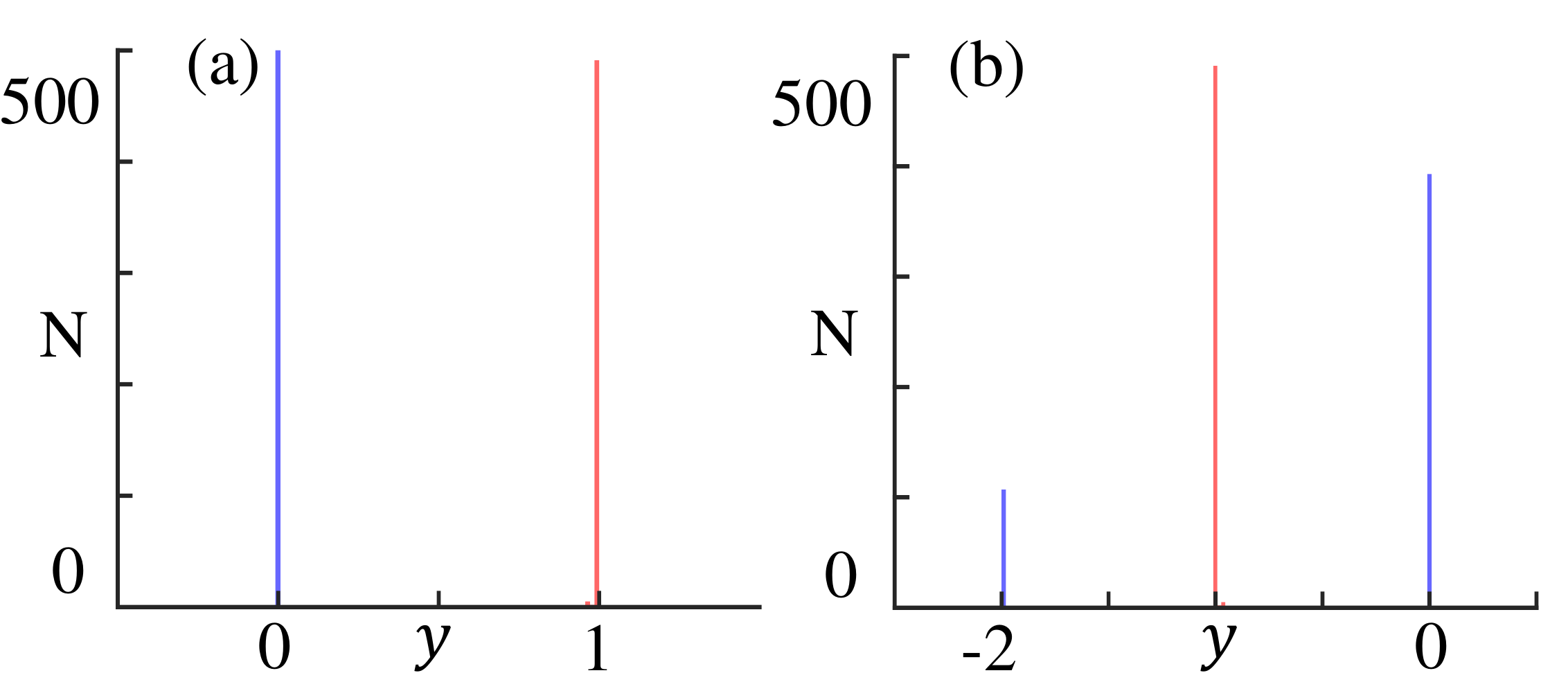}
    \caption{Topological classification of 1d Bloch Hamiltonians with particle-hole symmetry: The network's a) final output and b) intermediate output obtained before applying the predefined (mod 2) operation. The test data associated with the trivial reference (randomly chosen) parent is depicted in blue (red). }
    \label{fig:D_topo_numbers_test}
\end{figure}
We then trained a NN with a layout described in Sec.~\ref{sec:Train} and illustrated in Fig.~\ref{fig:1D_network}. The convolution layers of the net were chosen to have 128 filters for the first layer with (2, 2) receptive field, and  64, 32, 1 filters for the other ones with (1, 1) receptive field. The same as in Sec.~\ref{sec:1D_1}, the network was successfully trained without any modulo layer. In total there are $43 \, 366$ trainable parameters. The network was trained on $2 \cdot 9500$ states and tested on $2 \cdot 500$ states employing absolute mean error cost and Adam optimizer. The training was done over $10^3$ epochs with learning rate $\eta = 10^{-3}$. Before every epoch we performed a uniform symmetry-respecting rotation of each training state and by this effectively augmented the dataset. 

The outcome of our protocol is depicted in Fig.~\ref{fig:IS_topo_numbers_test}. The network has learned to differentiate between two ensembles of topologically equivalent children with precision $100\%$, Fig.~\ref{fig:IS_topo_numbers_test}a. By looking at the network's intermediate output we could also retrieve the local quantity corresponding to the learned topological index, as shown in Fig.~\ref{fig:IS_topo_numbers_test}b - e. Strikingly, the network learned the importance of high-symmetric $k=0, \pi$ and simply avoids all other momentum points. Important to note that this property was not built-in into the network by hand and it highlights the flexibility of our NN layout, Figs.~\ref{fig:1D_network}: The last dense layer in the network allows it to find important points, lines and/or surfaces in momentum space without any external supervision.

\subsection{Multiband 1D insulators with particle-hole symmetry~\label{sec:1D_3}}

For our last example we implement the protocol for 1d topological classification in symmetry class D composed of systems exhibiting a particle-hole symmetry. It is well known that the topological number distinguishing the phases within this symmetry class is a $Z_2$ number. The topologically nontrivial phase within this symmetry class exhibits very rich physics and opens up a road for realization of robust Majorana end modes~\cite{Lutchyn2018}.

In first quantization the particle-hole symmetry is antiunitary anticommuting symmetry explicitly described by relation $H(k) = -U_C^\dagger \, H^*(-k) \, U_C$ for some unitary operator $U_C$. Here we use the conventional representation of particle-hole symmetry that is basic within the BCS theory of superconductivity: We take $U_C = I \otimes \tau_x$ with a Pauli matrix $\tau_x$ and identity $I$. To symmetrize Hermitian matrices~$A$ and Bloch Hamiltonians $H(k)$ the following transformations are performed:
\begin{align}
    A &\rightarrow A - U_{C} A^* U_{C}^\dagger \\
    H(k)&\rightarrow H(k) - U_{C} H^*(-k) U_{C}^\dagger. 
    \label{eq:D}
\end{align}

By applying the gradient descent algorithm from Sec.~\ref{sec:Prerequisites}, it was found that the generated symmetrized matrices $A$ clustered into two blocks of matrices and we uniformly picked matrices from them for producing trivial atomic-limit Bloch Hamiltonians respecting the particle-hole symmetry. These two blocks of trivial Hamiltonians would in the Majorana basis correspond to the Pfaffian at $k=0$ and $k=\pi$ being $\pm 1$\cite{RevModPhys.88.035005, PhysRevB.88.075419}. In total there were  $2\cdot 10^3$ matrices generated. All the details on other aspects of the data generation are provided in Sec.~\ref{sec:1D_0}.

Again, in the training stage we employed a neural net layout shown in Fig.~\ref{fig:1D_network} with convolution layers of 256, 128, 64, 1 filters. The first convolution layer operates with (2, 2) receptive field and all other ones with (1, 1) receptive field. There are $107 \, 110$ adjustable parameters in total. Without a modulo layer the network has failed to adjust to separate the classes of the training data, but with the modulo layer depicted in Fig.~\ref{fig:1D_network} it has succeeded. We trained the network on $2 \cdot 9500$ and tested on $2 \cdot 500$ samples using Adam optimizer and absolute mean error cost function. The training was done over $10^3$ epochs with learning rate $\eta = 10^{-4}$. The same as in all previous cases we effectively augmented the training dataset by doing a uniform symmetry-respecting rotation of every Bloch Hamiltonian before each epoch.

\begin{figure}[t!] \centering
    \includegraphics[width=8.5cm,angle=0]{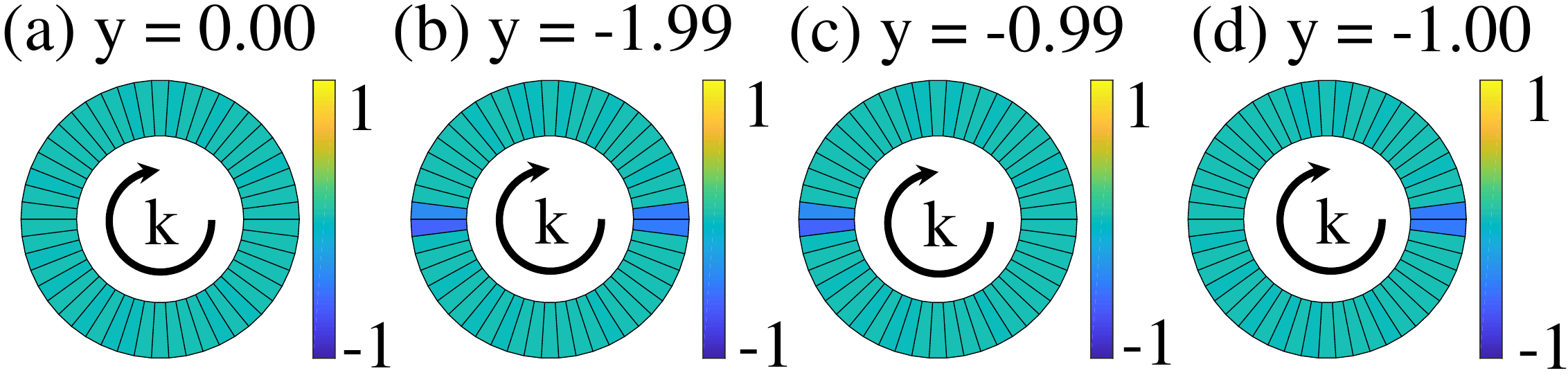}
    \caption{The learned local quantities ($y$) corresponding to particle-hole symmetric systems illustrated on four different examples, showing the (coarse grained) momentum resolved output before performing the modulo operation.}
    \label{fig:D_local}
\end{figure}

The network's output evaluated on a test dataset is shown in Fig.~\ref{fig:D_topo_numbers_test}: Two ensembles of topologically equivalent children are successfully differentiated by the network with high precision, Fig.~\ref{fig:D_topo_numbers_test}a. In Fig.~\ref{fig:D_topo_numbers_test}b we show the intermediate output produced by the network before applying the mod 2 operation: Interestingly, we see the data corresponding to topologically trivial ensemble clustered around two different values, $-2$ and $0$, confirming the $Z_2$ nature of the topological index. In Fig.~\ref{fig:D_local} four examples of test states show the local quantity corresponding to the learned topological number. Again, the network has adjusted to look at only high-symmetric $k=0, \pi$ and neglect other momentum points.  

These results demonstrate that the network without explicit guidance learns to focus on the pertinent information. In fact, it is known that these Hamiltonians can be classified by product of the sign of the Pfaffian at the high symmetry points $k=0$ and $\pi$. Without extending the datasets with different sectors of trivial atomic bands the network would only learn to pick up one of the high symmetry points, and it would fail to classify all Hamiltonians properly. One might train two networks to learn to separately classify the two symmetry points, but the corresponding topological indices will be a composition of topologically irrelevant and relevant invariants. Another alternative to the mod layer for this problem would be to use an additional dense layer with non-linear activation functions to effectuate an xor classification of the output from the last convolution layer. However, the drawback of this is that it would allow for a drift of the last feature map output, with a corresponding loss of the interpretability of the results.

\section{Conclusion and outlook}
We present a novel protocol for using NNs to learn topological indices of Bloch Hamiltonians, extending previous work on `topological data augmentation’. The protocol is characterized by the following features: i) It is unsupervised, i.e. training data is generated by randomly deforming parents states while ensuring the topological integrity of the data sets, without using any prior knowledge or specific models except a specific representation of the symmetry class to constrain the data. ii) The NNs are specifically designed to be interpretable, in the sense that the single feature map of the last convolution layer gives momentum resolved information about the learned quantities. In this way the network can for example learn to single out relevant high symmetry momentum points. iii) By extending the training data samples with a complete set of symmetry respecting  atomic insulators (specifically here, stacking 4 bands on 4 bands) the network learns to pick up only relevant topological invariants that do not discriminate between topologically disconnected but trivial atomic-limit Hamiltonians. 

The protocol takes a next step towards the goal of using machine learning and NNs to identify unknown topological invariants that go beyond the already well established non-interacting and translationally invariant systems with spatial or internal symmetries. {Areas to explore could be periodically-driven (Floquet) topological phases \cite{PhysRevB.82.235114, Lindner2011}, non-Hermitian topological matter \cite{PhysRevX.8.031079, PhysRevX.9.041015}, interacting systems \cite{PhysRevB.83.075103}, and others.}  The NNs used in this work are very small compared to state of the art deep learning networks which opens up for extending the protocol to systems with less symmetry (e.g.\ disordered systems) that would require dense matrix input which is not in a block diagonal form. A nice ingredient of the network structure for the examples presented in this work and in Ref. \cite{balabanov1} is the interpretability that follows from the construction where the network has local operations in momentum space, until the very last layers. It remains to be explored if and how this can be extended to problems where local measures may not be sufficient. Nevertheless, the interpretability is a bonus which is not strictly necessary for the general topological data augmentation procedure and network classification. \\


\begin{acknowledgments}
This work was supported by the Swedish Research Council through Grant No. 621-2014-5972. 

\end{acknowledgments}



\bibliography{refs_PRR2}

\begin{thebibliography}{63}%
\makeatletter
\providecommand \@ifxundefined [1]{%
 \@ifx{#1\undefined}
}%
\providecommand \@ifnum [1]{%
 \ifnum #1\expandafter \@firstoftwo
 \else \expandafter \@secondoftwo
 \fi
}%
\providecommand \@ifx [1]{%
 \ifx #1\expandafter \@firstoftwo
 \else \expandafter \@secondoftwo
 \fi
}%
\providecommand \natexlab [1]{#1}%
\providecommand \enquote  [1]{``#1''}%
\providecommand \bibnamefont  [1]{#1}%
\providecommand \bibfnamefont [1]{#1}%
\providecommand \citenamefont [1]{#1}%
\providecommand \href@noop [0]{\@secondoftwo}%
\providecommand \href [0]{\begingroup \@sanitize@url \@href}%
\providecommand \@href[1]{\@@startlink{#1}\@@href}%
\providecommand \@@href[1]{\endgroup#1\@@endlink}%
\providecommand \@sanitize@url [0]{\catcode `\\12\catcode `\$12\catcode
  `\&12\catcode `\#12\catcode `\^12\catcode `\_12\catcode `\%12\relax}%
\providecommand \@@startlink[1]{}%
\providecommand \@@endlink[0]{}%
\providecommand \url  [0]{\begingroup\@sanitize@url \@url }%
\providecommand \@url [1]{\endgroup\@href {#1}{\urlprefix }}%
\providecommand \urlprefix  [0]{URL }%
\providecommand \Eprint [0]{\href }%
\providecommand \doibase [0]{http://dx.doi.org/}%
\providecommand \selectlanguage [0]{\@gobble}%
\providecommand \bibinfo  [0]{\@secondoftwo}%
\providecommand \bibfield  [0]{\@secondoftwo}%
\providecommand \translation [1]{[#1]}%
\providecommand \BibitemOpen [0]{}%
\providecommand \bibitemStop [0]{}%
\providecommand \bibitemNoStop [0]{.\EOS\space}%
\providecommand \EOS [0]{\spacefactor3000\relax}%
\providecommand \BibitemShut  [1]{\csname bibitem#1\endcsname}%
\let\auto@bib@innerbib\@empty
\bibitem [{\citenamefont {Balabanov}\ and\ \citenamefont
  {Granath}(2020)}]{balabanov1}%
  \BibitemOpen
  \bibfield  {author} {\bibinfo {author} {\bibfnamefont {Oleksandr}\
  \bibnamefont {Balabanov}}\ and\ \bibinfo {author} {\bibfnamefont {Mats}\
  \bibnamefont {Granath}},\ }\bibfield  {title} {\enquote {\bibinfo {title}
  {Unsupervised learning using topological data augmentation},}\ }\href
  {\doibase 10.1103/PhysRevResearch.2.013354} {\bibfield  {journal} {\bibinfo
  {journal} {Phys. Rev. Research}\ }\textbf {\bibinfo {volume} {2}},\ \bibinfo
  {pages} {013354} (\bibinfo {year} {2020})}\BibitemShut {NoStop}%
\bibitem [{\citenamefont {Hasan}\ and\ \citenamefont
  {Kane}(2010)}]{RevModPhys.82.3045}%
  \BibitemOpen
  \bibfield  {author} {\bibinfo {author} {\bibfnamefont {M.~Z.}\ \bibnamefont
  {Hasan}}\ and\ \bibinfo {author} {\bibfnamefont {C.~L.}\ \bibnamefont
  {Kane}},\ }\bibfield  {title} {\enquote {\bibinfo {title} {Colloquium:
  Topological insulators},}\ }\href {\doibase 10.1103/RevModPhys.82.3045}
  {\bibfield  {journal} {\bibinfo  {journal} {Rev. Mod. Phys.}\ }\textbf
  {\bibinfo {volume} {82}},\ \bibinfo {pages} {3045--3067} (\bibinfo {year}
  {2010})}\BibitemShut {NoStop}%
\bibitem [{\citenamefont {Qi}\ and\ \citenamefont
  {Zhang}(2011)}]{RevModPhys.83.1057}%
  \BibitemOpen
  \bibfield  {author} {\bibinfo {author} {\bibfnamefont {Xiao-Liang}\
  \bibnamefont {Qi}}\ and\ \bibinfo {author} {\bibfnamefont {Shou-Cheng}\
  \bibnamefont {Zhang}},\ }\bibfield  {title} {\enquote {\bibinfo {title}
  {Topological insulators and superconductors},}\ }\href {\doibase
  10.1103/RevModPhys.83.1057} {\bibfield  {journal} {\bibinfo  {journal} {Rev.
  Mod. Phys.}\ }\textbf {\bibinfo {volume} {83}},\ \bibinfo {pages}
  {1057--1110} (\bibinfo {year} {2011})}\BibitemShut {NoStop}%
\bibitem [{\citenamefont {Chiu}\ \emph {et~al.}(2016)\citenamefont {Chiu},
  \citenamefont {Teo}, \citenamefont {Schnyder},\ and\ \citenamefont
  {Ryu}}]{RevModPhys.88.035005}%
  \BibitemOpen
  \bibfield  {author} {\bibinfo {author} {\bibfnamefont {Ching-Kai}\
  \bibnamefont {Chiu}}, \bibinfo {author} {\bibfnamefont {Jeffrey C.~Y.}\
  \bibnamefont {Teo}}, \bibinfo {author} {\bibfnamefont {Andreas~P.}\
  \bibnamefont {Schnyder}}, \ and\ \bibinfo {author} {\bibfnamefont {Shinsei}\
  \bibnamefont {Ryu}},\ }\bibfield  {title} {\enquote {\bibinfo {title}
  {Classification of topological quantum matter with symmetries},}\ }\href
  {\doibase 10.1103/RevModPhys.88.035005} {\bibfield  {journal} {\bibinfo
  {journal} {Rev. Mod. Phys.}\ }\textbf {\bibinfo {volume} {88}},\ \bibinfo
  {pages} {035005} (\bibinfo {year} {2016})}\BibitemShut {NoStop}%
\bibitem [{\citenamefont {Dunjko}\ and\ \citenamefont
  {Briegel}(2018)}]{dunjko2018machine}%
  \BibitemOpen
  \bibfield  {author} {\bibinfo {author} {\bibfnamefont {Vedran}\ \bibnamefont
  {Dunjko}}\ and\ \bibinfo {author} {\bibfnamefont {Hans~J}\ \bibnamefont
  {Briegel}},\ }\bibfield  {title} {\enquote {\bibinfo {title} {Machine
  learning \& artificial intelligence in the quantum domain: a review of recent
  progress},}\ }\href {\doibase 10.1088/1361-6633/aab406} {\bibfield  {journal}
  {\bibinfo  {journal} {Reports on Progress in Physics}\ }\textbf {\bibinfo
  {volume} {81}},\ \bibinfo {pages} {074001} (\bibinfo {year}
  {2018})}\BibitemShut {NoStop}%
\bibitem [{\citenamefont {Carrasquilla}(2020)}]{carrasquilla2020machine}%
  \BibitemOpen
  \bibfield  {author} {\bibinfo {author} {\bibfnamefont {Juan}\ \bibnamefont
  {Carrasquilla}},\ }\bibfield  {title} {\enquote {\bibinfo {title} {Machine
  learning for quantum matter},}\ }\href {https://arxiv.org/abs/2003.11040}
  {\bibfield  {journal} {\bibinfo  {journal} {arXiv:2003.11040}\ } (\bibinfo
  {year} {2020})}\BibitemShut {NoStop}%
\bibitem [{\citenamefont {Carleo}\ and\ \citenamefont
  {Troyer}(2017)}]{carleo2017solving}%
  \BibitemOpen
  \bibfield  {author} {\bibinfo {author} {\bibfnamefont {Giuseppe}\
  \bibnamefont {Carleo}}\ and\ \bibinfo {author} {\bibfnamefont {Matthias}\
  \bibnamefont {Troyer}},\ }\bibfield  {title} {\enquote {\bibinfo {title}
  {Solving the quantum many-body problem with artificial neural networks},}\
  }\href {\doibase 10.1126/science.aag2302} {\bibfield  {journal} {\bibinfo
  {journal} {Science}\ }\textbf {\bibinfo {volume} {355}},\ \bibinfo {pages}
  {602--606} (\bibinfo {year} {2017})}\BibitemShut {NoStop}%
\bibitem [{\citenamefont {Gao}\ and\ \citenamefont {Duan}(2017)}]{Gao2017}%
  \BibitemOpen
  \bibfield  {author} {\bibinfo {author} {\bibfnamefont {Xun}\ \bibnamefont
  {Gao}}\ and\ \bibinfo {author} {\bibfnamefont {Lu-Ming}\ \bibnamefont
  {Duan}},\ }\bibfield  {title} {\enquote {\bibinfo {title} {Efficient
  representation of quantum many-body states with deep neural networks},}\
  }\href {\doibase 10.1038/s41467-017-00705-2} {\bibfield  {journal} {\bibinfo
  {journal} {Nature Communications}\ }\textbf {\bibinfo {volume} {8}},\
  \bibinfo {pages} {662} (\bibinfo {year} {2017})}\BibitemShut {NoStop}%
\bibitem [{\citenamefont {Deng}\ \emph
  {et~al.}(2017{\natexlab{a}})\citenamefont {Deng}, \citenamefont {Li},\ and\
  \citenamefont {Das~Sarma}}]{PhysRevX.7.021021}%
  \BibitemOpen
  \bibfield  {author} {\bibinfo {author} {\bibfnamefont {Dong-Ling}\
  \bibnamefont {Deng}}, \bibinfo {author} {\bibfnamefont {Xiaopeng}\
  \bibnamefont {Li}}, \ and\ \bibinfo {author} {\bibfnamefont {S.}~\bibnamefont
  {Das~Sarma}},\ }\bibfield  {title} {\enquote {\bibinfo {title} {Quantum
  entanglement in neural network states},}\ }\href {\doibase
  10.1103/PhysRevX.7.021021} {\bibfield  {journal} {\bibinfo  {journal} {Phys.
  Rev. X}\ }\textbf {\bibinfo {volume} {7}},\ \bibinfo {pages} {021021}
  (\bibinfo {year} {2017}{\natexlab{a}})}\BibitemShut {NoStop}%
\bibitem [{\citenamefont {Deng}\ \emph
  {et~al.}(2017{\natexlab{b}})\citenamefont {Deng}, \citenamefont {Li},\ and\
  \citenamefont {Das~Sarma}}]{PhysRevB.96.195145}%
  \BibitemOpen
  \bibfield  {author} {\bibinfo {author} {\bibfnamefont {Dong-Ling}\
  \bibnamefont {Deng}}, \bibinfo {author} {\bibfnamefont {Xiaopeng}\
  \bibnamefont {Li}}, \ and\ \bibinfo {author} {\bibfnamefont {S.}~\bibnamefont
  {Das~Sarma}},\ }\bibfield  {title} {\enquote {\bibinfo {title} {Machine
  learning topological states},}\ }\href {\doibase 10.1103/PhysRevB.96.195145}
  {\bibfield  {journal} {\bibinfo  {journal} {Phys. Rev. B}\ }\textbf {\bibinfo
  {volume} {96}},\ \bibinfo {pages} {195145} (\bibinfo {year}
  {2017}{\natexlab{b}})}\BibitemShut {NoStop}%
\bibitem [{\citenamefont {Nomura}\ \emph {et~al.}(2017)\citenamefont {Nomura},
  \citenamefont {Darmawan}, \citenamefont {Yamaji},\ and\ \citenamefont
  {Imada}}]{PhysRevB.96.205152}%
  \BibitemOpen
  \bibfield  {author} {\bibinfo {author} {\bibfnamefont {Yusuke}\ \bibnamefont
  {Nomura}}, \bibinfo {author} {\bibfnamefont {Andrew~S.}\ \bibnamefont
  {Darmawan}}, \bibinfo {author} {\bibfnamefont {Youhei}\ \bibnamefont
  {Yamaji}}, \ and\ \bibinfo {author} {\bibfnamefont {Masatoshi}\ \bibnamefont
  {Imada}},\ }\bibfield  {title} {\enquote {\bibinfo {title} {Restricted
  boltzmann machine learning for solving strongly correlated quantum
  systems},}\ }\href {\doibase 10.1103/PhysRevB.96.205152} {\bibfield
  {journal} {\bibinfo  {journal} {Phys. Rev. B}\ }\textbf {\bibinfo {volume}
  {96}},\ \bibinfo {pages} {205152} (\bibinfo {year} {2017})}\BibitemShut
  {NoStop}%
\bibitem [{\citenamefont {Kaubruegger}\ \emph {et~al.}(2018)\citenamefont
  {Kaubruegger}, \citenamefont {Pastori},\ and\ \citenamefont
  {Budich}}]{PhysRevB.97.195136}%
  \BibitemOpen
  \bibfield  {author} {\bibinfo {author} {\bibfnamefont {Raphael}\ \bibnamefont
  {Kaubruegger}}, \bibinfo {author} {\bibfnamefont {Lorenzo}\ \bibnamefont
  {Pastori}}, \ and\ \bibinfo {author} {\bibfnamefont {Jan~Carl}\ \bibnamefont
  {Budich}},\ }\bibfield  {title} {\enquote {\bibinfo {title} {Chiral
  topological phases from artificial neural networks},}\ }\href {\doibase
  10.1103/PhysRevB.97.195136} {\bibfield  {journal} {\bibinfo  {journal} {Phys.
  Rev. B}\ }\textbf {\bibinfo {volume} {97}},\ \bibinfo {pages} {195136}
  (\bibinfo {year} {2018})}\BibitemShut {NoStop}%
\bibitem [{\citenamefont {Pastori}\ \emph {et~al.}(2019)\citenamefont
  {Pastori}, \citenamefont {Kaubruegger},\ and\ \citenamefont
  {Budich}}]{PhysRevB.99.165123}%
  \BibitemOpen
  \bibfield  {author} {\bibinfo {author} {\bibfnamefont {Lorenzo}\ \bibnamefont
  {Pastori}}, \bibinfo {author} {\bibfnamefont {Raphael}\ \bibnamefont
  {Kaubruegger}}, \ and\ \bibinfo {author} {\bibfnamefont {Jan~Carl}\
  \bibnamefont {Budich}},\ }\bibfield  {title} {\enquote {\bibinfo {title}
  {Generalized transfer matrix states from artificial neural networks},}\
  }\href {\doibase 10.1103/PhysRevB.99.165123} {\bibfield  {journal} {\bibinfo
  {journal} {Phys. Rev. B}\ }\textbf {\bibinfo {volume} {99}},\ \bibinfo
  {pages} {165123} (\bibinfo {year} {2019})}\BibitemShut {NoStop}%
\bibitem [{\citenamefont {Melko}\ \emph {et~al.}(2019)\citenamefont {Melko},
  \citenamefont {Carleo}, \citenamefont {Carrasquilla},\ and\ \citenamefont
  {Cirac}}]{melko2019restricted}%
  \BibitemOpen
  \bibfield  {author} {\bibinfo {author} {\bibfnamefont {Roger~G}\ \bibnamefont
  {Melko}}, \bibinfo {author} {\bibfnamefont {Giuseppe}\ \bibnamefont
  {Carleo}}, \bibinfo {author} {\bibfnamefont {Juan}\ \bibnamefont
  {Carrasquilla}}, \ and\ \bibinfo {author} {\bibfnamefont {J~Ignacio}\
  \bibnamefont {Cirac}},\ }\bibfield  {title} {\enquote {\bibinfo {title}
  {Restricted boltzmann machines in quantum physics},}\ }\href {\doibase
  10.1038/s41567-019-0545-1} {\bibfield  {journal} {\bibinfo  {journal} {Nature
  Physics}\ }\textbf {\bibinfo {volume} {15}},\ \bibinfo {pages} {887--892}
  (\bibinfo {year} {2019})}\BibitemShut {NoStop}%
\bibitem [{\citenamefont {Glasser}\ \emph {et~al.}(2018)\citenamefont
  {Glasser}, \citenamefont {Pancotti}, \citenamefont {August}, \citenamefont
  {Rodriguez},\ and\ \citenamefont {Cirac}}]{PhysRevX.8.011006}%
  \BibitemOpen
  \bibfield  {author} {\bibinfo {author} {\bibfnamefont {Ivan}\ \bibnamefont
  {Glasser}}, \bibinfo {author} {\bibfnamefont {Nicola}\ \bibnamefont
  {Pancotti}}, \bibinfo {author} {\bibfnamefont {Moritz}\ \bibnamefont
  {August}}, \bibinfo {author} {\bibfnamefont {Ivan~D.}\ \bibnamefont
  {Rodriguez}}, \ and\ \bibinfo {author} {\bibfnamefont {J.~Ignacio}\
  \bibnamefont {Cirac}},\ }\bibfield  {title} {\enquote {\bibinfo {title}
  {Neural-network quantum states, string-bond states, and chiral topological
  states},}\ }\href {\doibase 10.1103/PhysRevX.8.011006} {\bibfield  {journal}
  {\bibinfo  {journal} {Phys. Rev. X}\ }\textbf {\bibinfo {volume} {8}},\
  \bibinfo {pages} {011006} (\bibinfo {year} {2018})}\BibitemShut {NoStop}%
\bibitem [{\citenamefont {Choo}\ \emph {et~al.}(2018)\citenamefont {Choo},
  \citenamefont {Carleo}, \citenamefont {Regnault},\ and\ \citenamefont
  {Neupert}}]{PhysRevLett.121.167204}%
  \BibitemOpen
  \bibfield  {author} {\bibinfo {author} {\bibfnamefont {Kenny}\ \bibnamefont
  {Choo}}, \bibinfo {author} {\bibfnamefont {Giuseppe}\ \bibnamefont {Carleo}},
  \bibinfo {author} {\bibfnamefont {Nicolas}\ \bibnamefont {Regnault}}, \ and\
  \bibinfo {author} {\bibfnamefont {Titus}\ \bibnamefont {Neupert}},\
  }\bibfield  {title} {\enquote {\bibinfo {title} {Symmetries and many-body
  excitations with neural-network quantum states},}\ }\href {\doibase
  10.1103/PhysRevLett.121.167204} {\bibfield  {journal} {\bibinfo  {journal}
  {Phys. Rev. Lett.}\ }\textbf {\bibinfo {volume} {121}},\ \bibinfo {pages}
  {167204} (\bibinfo {year} {2018})}\BibitemShut {NoStop}%
\bibitem [{\citenamefont {Zhang}\ \emph {et~al.}(2018)\citenamefont {Zhang},
  \citenamefont {Shen},\ and\ \citenamefont {Zhai}}]{PhysRevLett.120.066401}%
  \BibitemOpen
  \bibfield  {author} {\bibinfo {author} {\bibfnamefont {Pengfei}\ \bibnamefont
  {Zhang}}, \bibinfo {author} {\bibfnamefont {Huitao}\ \bibnamefont {Shen}}, \
  and\ \bibinfo {author} {\bibfnamefont {Hui}\ \bibnamefont {Zhai}},\
  }\bibfield  {title} {\enquote {\bibinfo {title} {Machine learning topological
  invariants with neural networks},}\ }\href {\doibase
  10.1103/PhysRevLett.120.066401} {\bibfield  {journal} {\bibinfo  {journal}
  {Phys. Rev. Lett.}\ }\textbf {\bibinfo {volume} {120}},\ \bibinfo {pages}
  {066401} (\bibinfo {year} {2018})}\BibitemShut {NoStop}%
\bibitem [{\citenamefont {Sun}\ \emph {et~al.}(2018)\citenamefont {Sun},
  \citenamefont {Yi}, \citenamefont {Zhang}, \citenamefont {Shen},\ and\
  \citenamefont {Zhai}}]{PhysRevB.98.085402}%
  \BibitemOpen
  \bibfield  {author} {\bibinfo {author} {\bibfnamefont {Ning}\ \bibnamefont
  {Sun}}, \bibinfo {author} {\bibfnamefont {Jinmin}\ \bibnamefont {Yi}},
  \bibinfo {author} {\bibfnamefont {Pengfei}\ \bibnamefont {Zhang}}, \bibinfo
  {author} {\bibfnamefont {Huitao}\ \bibnamefont {Shen}}, \ and\ \bibinfo
  {author} {\bibfnamefont {Hui}\ \bibnamefont {Zhai}},\ }\bibfield  {title}
  {\enquote {\bibinfo {title} {Deep learning topological invariants of band
  insulators},}\ }\href {\doibase 10.1103/PhysRevB.98.085402} {\bibfield
  {journal} {\bibinfo  {journal} {Phys. Rev. B}\ }\textbf {\bibinfo {volume}
  {98}},\ \bibinfo {pages} {085402} (\bibinfo {year} {2018})}\BibitemShut
  {NoStop}%
\bibitem [{\citenamefont {Casert}\ \emph {et~al.}(2019)\citenamefont {Casert},
  \citenamefont {Vieijra}, \citenamefont {Nys},\ and\ \citenamefont
  {Ryckebusch}}]{PhysRevE.99.023304}%
  \BibitemOpen
  \bibfield  {author} {\bibinfo {author} {\bibfnamefont {C.}~\bibnamefont
  {Casert}}, \bibinfo {author} {\bibfnamefont {T.}~\bibnamefont {Vieijra}},
  \bibinfo {author} {\bibfnamefont {J.}~\bibnamefont {Nys}}, \ and\ \bibinfo
  {author} {\bibfnamefont {J.}~\bibnamefont {Ryckebusch}},\ }\bibfield  {title}
  {\enquote {\bibinfo {title} {Interpretable machine learning for inferring the
  phase boundaries in a nonequilibrium system},}\ }\href {\doibase
  10.1103/PhysRevE.99.023304} {\bibfield  {journal} {\bibinfo  {journal} {Phys.
  Rev. E}\ }\textbf {\bibinfo {volume} {99}},\ \bibinfo {pages} {023304}
  (\bibinfo {year} {2019})}\BibitemShut {NoStop}%
\bibitem [{\citenamefont {Carvalho}\ \emph {et~al.}(2018)\citenamefont
  {Carvalho}, \citenamefont {Garc\'{\i}a-Mart\'{\i}nez}, \citenamefont {Lado},\
  and\ \citenamefont {Fern\'andez-Rossier}}]{PhysRevB.97.115453}%
  \BibitemOpen
  \bibfield  {author} {\bibinfo {author} {\bibfnamefont {D.}~\bibnamefont
  {Carvalho}}, \bibinfo {author} {\bibfnamefont {N.~A.}\ \bibnamefont
  {Garc\'{\i}a-Mart\'{\i}nez}}, \bibinfo {author} {\bibfnamefont {J.~L.}\
  \bibnamefont {Lado}}, \ and\ \bibinfo {author} {\bibfnamefont
  {J.}~\bibnamefont {Fern\'andez-Rossier}},\ }\bibfield  {title} {\enquote
  {\bibinfo {title} {Real-space mapping of topological invariants using
  artificial neural networks},}\ }\href {\doibase 10.1103/PhysRevB.97.115453}
  {\bibfield  {journal} {\bibinfo  {journal} {Phys. Rev. B}\ }\textbf {\bibinfo
  {volume} {97}},\ \bibinfo {pages} {115453} (\bibinfo {year}
  {2018})}\BibitemShut {NoStop}%
\bibitem [{\citenamefont {Ming}\ \emph {et~al.}(2019)\citenamefont {Ming},
  \citenamefont {Lin}, \citenamefont {Bartlett},\ and\ \citenamefont
  {Zhang}}]{Ming2019}%
  \BibitemOpen
  \bibfield  {author} {\bibinfo {author} {\bibfnamefont {Yurui}\ \bibnamefont
  {Ming}}, \bibinfo {author} {\bibfnamefont {Chin-Teng}\ \bibnamefont {Lin}},
  \bibinfo {author} {\bibfnamefont {Stephen~D.}\ \bibnamefont {Bartlett}}, \
  and\ \bibinfo {author} {\bibfnamefont {Wei-Wei}\ \bibnamefont {Zhang}},\
  }\bibfield  {title} {\enquote {\bibinfo {title} {Quantum topology
  identification with deep neural networks and quantum walks},}\ }\href
  {\doibase 10.1038/s41524-019-0224-x} {\bibfield  {journal} {\bibinfo
  {journal} {npj Computational Materials}\ }\textbf {\bibinfo {volume} {5}},\
  \bibinfo {pages} {88} (\bibinfo {year} {2019})}\BibitemShut {NoStop}%
\bibitem [{\citenamefont {Greplova}\ \emph {et~al.}(2020)\citenamefont
  {Greplova}, \citenamefont {Valenti}, \citenamefont {Boschung}, \citenamefont
  {Schäfer}, \citenamefont {Lörch},\ and\ \citenamefont
  {Huber}}]{Greplova_2020}%
  \BibitemOpen
  \bibfield  {author} {\bibinfo {author} {\bibfnamefont {Eliska}\ \bibnamefont
  {Greplova}}, \bibinfo {author} {\bibfnamefont {Agnes}\ \bibnamefont
  {Valenti}}, \bibinfo {author} {\bibfnamefont {Gregor}\ \bibnamefont
  {Boschung}}, \bibinfo {author} {\bibfnamefont {Frank}\ \bibnamefont
  {Schäfer}}, \bibinfo {author} {\bibfnamefont {Niels}\ \bibnamefont
  {Lörch}}, \ and\ \bibinfo {author} {\bibfnamefont {Sebastian~D}\
  \bibnamefont {Huber}},\ }\bibfield  {title} {\enquote {\bibinfo {title}
  {Unsupervised identification of topological phase transitions using
  predictive models},}\ }\href {\doibase 10.1088/1367-2630/ab7771} {\bibfield
  {journal} {\bibinfo  {journal} {New Journal of Physics}\ }\textbf {\bibinfo
  {volume} {22}},\ \bibinfo {pages} {045003} (\bibinfo {year}
  {2020})}\BibitemShut {NoStop}%
\bibitem [{\citenamefont {{Tsai}}\ \emph {et~al.}(2019)\citenamefont {{Tsai}},
  \citenamefont {{Yu}}, \citenamefont {{Hsu}},\ and\ \citenamefont
  {{Chung}}}]{2019arXiv190904784T}%
  \BibitemOpen
  \bibfield  {author} {\bibinfo {author} {\bibfnamefont {Yuan-Hong}\
  \bibnamefont {{Tsai}}}, \bibinfo {author} {\bibfnamefont {Meng-Zhe}\
  \bibnamefont {{Yu}}}, \bibinfo {author} {\bibfnamefont {Yu-Hao}\ \bibnamefont
  {{Hsu}}}, \ and\ \bibinfo {author} {\bibfnamefont {Ming-Chiang}\ \bibnamefont
  {{Chung}}},\ }\bibfield  {title} {\enquote {\bibinfo {title} {{Deep learning
  of topological phase transitions from entanglement aspects}},}\ }\href
  {https://ui.adsabs.harvard.edu/abs/2019arXiv190904784T} {\bibfield  {journal}
  {\bibinfo  {journal} {arXiv:1909.04784}\ } (\bibinfo {year}
  {2019})}\BibitemShut {NoStop}%
\bibitem [{\citenamefont {{Caio}}\ \emph {et~al.}(2019)\citenamefont {{Caio}},
  \citenamefont {{Caccin}}, \citenamefont {{Baireuther}}, \citenamefont
  {{Hyart}},\ and\ \citenamefont {{Fruchart}}}]{2019arXiv190103346C}%
  \BibitemOpen
  \bibfield  {author} {\bibinfo {author} {\bibfnamefont {Marcello~D.}\
  \bibnamefont {{Caio}}}, \bibinfo {author} {\bibfnamefont {Marco}\
  \bibnamefont {{Caccin}}}, \bibinfo {author} {\bibfnamefont {Paul}\
  \bibnamefont {{Baireuther}}}, \bibinfo {author} {\bibfnamefont {Timo}\
  \bibnamefont {{Hyart}}}, \ and\ \bibinfo {author} {\bibfnamefont {Michel}\
  \bibnamefont {{Fruchart}}},\ }\bibfield  {title} {\enquote {\bibinfo {title}
  {{Machine learning assisted measurement of local topological invariants}},}\
  }\href {https://ui.adsabs.harvard.edu/abs/2019arXiv190103346C} {\bibfield
  {journal} {\bibinfo  {journal} {arXiv:1901.03346}\ } (\bibinfo {year}
  {2019})}\BibitemShut {NoStop}%
\bibitem [{\citenamefont {Zhang}\ \emph {et~al.}(2020)\citenamefont {Zhang},
  \citenamefont {Ginsparg},\ and\ \citenamefont
  {Kim}}]{PhysRevResearch.2.023283}%
  \BibitemOpen
  \bibfield  {author} {\bibinfo {author} {\bibfnamefont {Yi}~\bibnamefont
  {Zhang}}, \bibinfo {author} {\bibfnamefont {Paul}\ \bibnamefont {Ginsparg}},
  \ and\ \bibinfo {author} {\bibfnamefont {Eun-Ah}\ \bibnamefont {Kim}},\
  }\bibfield  {title} {\enquote {\bibinfo {title} {Interpreting machine
  learning of topological quantum phase transitions},}\ }\href {\doibase
  10.1103/PhysRevResearch.2.023283} {\bibfield  {journal} {\bibinfo  {journal}
  {Phys. Rev. Research}\ }\textbf {\bibinfo {volume} {2}},\ \bibinfo {pages}
  {023283} (\bibinfo {year} {2020})}\BibitemShut {NoStop}%
\bibitem [{\citenamefont {Carrasquilla}\ and\ \citenamefont
  {Melko}(2017)}]{Carrasquilla2017}%
  \BibitemOpen
  \bibfield  {author} {\bibinfo {author} {\bibfnamefont {Juan}\ \bibnamefont
  {Carrasquilla}}\ and\ \bibinfo {author} {\bibfnamefont {Roger~G.}\
  \bibnamefont {Melko}},\ }\bibfield  {title} {\enquote {\bibinfo {title}
  {Machine learning phases of matter},}\ }\href {\doibase 10.1038/nphys4035}
  {\bibfield  {journal} {\bibinfo  {journal} {Nature Physics}\ }\textbf
  {\bibinfo {volume} {13}},\ \bibinfo {pages} {431--434} (\bibinfo {year}
  {2017})}\BibitemShut {NoStop}%
\bibitem [{\citenamefont {Ch'ng}\ \emph {et~al.}(2017)\citenamefont {Ch'ng},
  \citenamefont {Carrasquilla}, \citenamefont {Melko},\ and\ \citenamefont
  {Khatami}}]{PhysRevX.7.031038}%
  \BibitemOpen
  \bibfield  {author} {\bibinfo {author} {\bibfnamefont {Kelvin}\ \bibnamefont
  {Ch'ng}}, \bibinfo {author} {\bibfnamefont {Juan}\ \bibnamefont
  {Carrasquilla}}, \bibinfo {author} {\bibfnamefont {Roger~G.}\ \bibnamefont
  {Melko}}, \ and\ \bibinfo {author} {\bibfnamefont {Ehsan}\ \bibnamefont
  {Khatami}},\ }\bibfield  {title} {\enquote {\bibinfo {title} {Machine
  learning phases of strongly correlated fermions},}\ }\href {\doibase
  10.1103/PhysRevX.7.031038} {\bibfield  {journal} {\bibinfo  {journal} {Phys.
  Rev. X}\ }\textbf {\bibinfo {volume} {7}},\ \bibinfo {pages} {031038}
  (\bibinfo {year} {2017})}\BibitemShut {NoStop}%
\bibitem [{\citenamefont {Broecker}\ \emph {et~al.}(2017)\citenamefont
  {Broecker}, \citenamefont {Carrasquilla}, \citenamefont {Melko},\ and\
  \citenamefont {Trebst}}]{Broecker2017}%
  \BibitemOpen
  \bibfield  {author} {\bibinfo {author} {\bibfnamefont {Peter}\ \bibnamefont
  {Broecker}}, \bibinfo {author} {\bibfnamefont {Juan}\ \bibnamefont
  {Carrasquilla}}, \bibinfo {author} {\bibfnamefont {Roger~G.}\ \bibnamefont
  {Melko}}, \ and\ \bibinfo {author} {\bibfnamefont {Simon}\ \bibnamefont
  {Trebst}},\ }\bibfield  {title} {\enquote {\bibinfo {title} {Machine learning
  quantum phases of matter beyond the fermion sign problem},}\ }\href {\doibase
  10.1038/s41598-017-09098-0} {\bibfield  {journal} {\bibinfo  {journal}
  {Scientific Reports}\ }\textbf {\bibinfo {volume} {7}},\ \bibinfo {pages}
  {8823} (\bibinfo {year} {2017})}\BibitemShut {NoStop}%
\bibitem [{\citenamefont {Beach}\ \emph {et~al.}(2018)\citenamefont {Beach},
  \citenamefont {Golubeva},\ and\ \citenamefont {Melko}}]{PhysRevB.97.045207}%
  \BibitemOpen
  \bibfield  {author} {\bibinfo {author} {\bibfnamefont {Matthew J.~S.}\
  \bibnamefont {Beach}}, \bibinfo {author} {\bibfnamefont {Anna}\ \bibnamefont
  {Golubeva}}, \ and\ \bibinfo {author} {\bibfnamefont {Roger~G.}\ \bibnamefont
  {Melko}},\ }\bibfield  {title} {\enquote {\bibinfo {title} {Machine learning
  vortices at the kosterlitz-thouless transition},}\ }\href {\doibase
  10.1103/PhysRevB.97.045207} {\bibfield  {journal} {\bibinfo  {journal} {Phys.
  Rev. B}\ }\textbf {\bibinfo {volume} {97}},\ \bibinfo {pages} {045207}
  (\bibinfo {year} {2018})}\BibitemShut {NoStop}%
\bibitem [{\citenamefont {Huembeli}\ \emph {et~al.}(2018)\citenamefont
  {Huembeli}, \citenamefont {Dauphin},\ and\ \citenamefont
  {Wittek}}]{PhysRevB.97.134109}%
  \BibitemOpen
  \bibfield  {author} {\bibinfo {author} {\bibfnamefont {Patrick}\ \bibnamefont
  {Huembeli}}, \bibinfo {author} {\bibfnamefont {Alexandre}\ \bibnamefont
  {Dauphin}}, \ and\ \bibinfo {author} {\bibfnamefont {Peter}\ \bibnamefont
  {Wittek}},\ }\bibfield  {title} {\enquote {\bibinfo {title} {Identifying
  quantum phase transitions with adversarial neural networks},}\ }\href
  {\doibase 10.1103/PhysRevB.97.134109} {\bibfield  {journal} {\bibinfo
  {journal} {Phys. Rev. B}\ }\textbf {\bibinfo {volume} {97}},\ \bibinfo
  {pages} {134109} (\bibinfo {year} {2018})}\BibitemShut {NoStop}%
\bibitem [{\citenamefont {Zhang}\ \emph {et~al.}(2019)\citenamefont {Zhang},
  \citenamefont {Liu},\ and\ \citenamefont {Wei}}]{PhysRevE.99.032142}%
  \BibitemOpen
  \bibfield  {author} {\bibinfo {author} {\bibfnamefont {Wanzhou}\ \bibnamefont
  {Zhang}}, \bibinfo {author} {\bibfnamefont {Jiayu}\ \bibnamefont {Liu}}, \
  and\ \bibinfo {author} {\bibfnamefont {Tzu-Chieh}\ \bibnamefont {Wei}},\
  }\bibfield  {title} {\enquote {\bibinfo {title} {Machine learning of phase
  transitions in the percolation and $xy$ models},}\ }\href {\doibase
  10.1103/PhysRevE.99.032142} {\bibfield  {journal} {\bibinfo  {journal} {Phys.
  Rev. E}\ }\textbf {\bibinfo {volume} {99}},\ \bibinfo {pages} {032142}
  (\bibinfo {year} {2019})}\BibitemShut {NoStop}%
\bibitem [{\citenamefont {Zhang}\ \emph {et~al.}(2017)\citenamefont {Zhang},
  \citenamefont {Melko},\ and\ \citenamefont {Kim}}]{PhysRevB.96.245119}%
  \BibitemOpen
  \bibfield  {author} {\bibinfo {author} {\bibfnamefont {Yi}~\bibnamefont
  {Zhang}}, \bibinfo {author} {\bibfnamefont {Roger~G.}\ \bibnamefont {Melko}},
  \ and\ \bibinfo {author} {\bibfnamefont {Eun-Ah}\ \bibnamefont {Kim}},\
  }\bibfield  {title} {\enquote {\bibinfo {title} {Machine learning $z_2$
  quantum spin liquids with quasiparticle statistics},}\ }\href {\doibase
  10.1103/PhysRevB.96.245119} {\bibfield  {journal} {\bibinfo  {journal} {Phys.
  Rev. B}\ }\textbf {\bibinfo {volume} {96}},\ \bibinfo {pages} {245119}
  (\bibinfo {year} {2017})}\BibitemShut {NoStop}%
\bibitem [{\citenamefont {Rem}\ \emph {et~al.}(2019)\citenamefont {Rem},
  \citenamefont {K{\"a}ming}, \citenamefont {Tarnowski}, \citenamefont
  {Asteria}, \citenamefont {Fl{\"a}schner}, \citenamefont {Becker},
  \citenamefont {Sengstock},\ and\ \citenamefont {Weitenberg}}]{Rem2019}%
  \BibitemOpen
  \bibfield  {author} {\bibinfo {author} {\bibfnamefont {Benno~S.}\
  \bibnamefont {Rem}}, \bibinfo {author} {\bibfnamefont {Niklas}\ \bibnamefont
  {K{\"a}ming}}, \bibinfo {author} {\bibfnamefont {Matthias}\ \bibnamefont
  {Tarnowski}}, \bibinfo {author} {\bibfnamefont {Luca}\ \bibnamefont
  {Asteria}}, \bibinfo {author} {\bibfnamefont {Nick}\ \bibnamefont
  {Fl{\"a}schner}}, \bibinfo {author} {\bibfnamefont {Christoph}\ \bibnamefont
  {Becker}}, \bibinfo {author} {\bibfnamefont {Klaus}\ \bibnamefont
  {Sengstock}}, \ and\ \bibinfo {author} {\bibfnamefont {Christof}\
  \bibnamefont {Weitenberg}},\ }\bibfield  {title} {\enquote {\bibinfo {title}
  {Identifying quantum phase transitions using artificial neural networks on
  experimental data},}\ }\href {\doibase 10.1038/s41567-019-0554-0} {\bibfield
  {journal} {\bibinfo  {journal} {Nature Physics}\ }\textbf {\bibinfo {volume}
  {15}},\ \bibinfo {pages} {917--920} (\bibinfo {year} {2019})}\BibitemShut
  {NoStop}%
\bibitem [{\citenamefont {Huembeli}\ \emph {et~al.}(2019)\citenamefont
  {Huembeli}, \citenamefont {Dauphin}, \citenamefont {Wittek},\ and\
  \citenamefont {Gogolin}}]{PhysRevB.99.104106}%
  \BibitemOpen
  \bibfield  {author} {\bibinfo {author} {\bibfnamefont {Patrick}\ \bibnamefont
  {Huembeli}}, \bibinfo {author} {\bibfnamefont {Alexandre}\ \bibnamefont
  {Dauphin}}, \bibinfo {author} {\bibfnamefont {Peter}\ \bibnamefont {Wittek}},
  \ and\ \bibinfo {author} {\bibfnamefont {Christian}\ \bibnamefont
  {Gogolin}},\ }\bibfield  {title} {\enquote {\bibinfo {title} {Automated
  discovery of characteristic features of phase transitions in many-body
  localization},}\ }\href {\doibase 10.1103/PhysRevB.99.104106} {\bibfield
  {journal} {\bibinfo  {journal} {Phys. Rev. B}\ }\textbf {\bibinfo {volume}
  {99}},\ \bibinfo {pages} {104106} (\bibinfo {year} {2019})}\BibitemShut
  {NoStop}%
\bibitem [{\citenamefont {{Goetz}}\ \emph {et~al.}(2019)\citenamefont
  {{Goetz}}, \citenamefont {{Zhang}},\ and\ \citenamefont
  {{Lawler}}}]{2019arXiv190111042G}%
  \BibitemOpen
  \bibfield  {author} {\bibinfo {author} {\bibfnamefont {Jeremy~B.}\
  \bibnamefont {{Goetz}}}, \bibinfo {author} {\bibfnamefont {Yi}~\bibnamefont
  {{Zhang}}}, \ and\ \bibinfo {author} {\bibfnamefont {Michael~J.}\
  \bibnamefont {{Lawler}}},\ }\bibfield  {title} {\enquote {\bibinfo {title}
  {{Detecting Nematic Order in STM/STS Data with Artificial Intelligence}},}\
  }\href {https://ui.adsabs.harvard.edu/abs/2019arXiv190111042G} {\bibfield
  {journal} {\bibinfo  {journal} {arXiv:1901.11042}\ } (\bibinfo {year}
  {2019})}\BibitemShut {NoStop}%
\bibitem [{\citenamefont {Van~Nieuwenburg}\ \emph {et~al.}(2017)\citenamefont
  {Van~Nieuwenburg}, \citenamefont {Liu},\ and\ \citenamefont
  {Huber}}]{van2017learning}%
  \BibitemOpen
  \bibfield  {author} {\bibinfo {author} {\bibfnamefont {Evert~PL}\
  \bibnamefont {Van~Nieuwenburg}}, \bibinfo {author} {\bibfnamefont {Ye-Hua}\
  \bibnamefont {Liu}}, \ and\ \bibinfo {author} {\bibfnamefont {Sebastian~D}\
  \bibnamefont {Huber}},\ }\bibfield  {title} {\enquote {\bibinfo {title}
  {Learning phase transitions by confusion},}\ }\href {\doibase
  10.1038/nphys4037} {\bibfield  {journal} {\bibinfo  {journal} {Nature
  Physics}\ }\textbf {\bibinfo {volume} {13}},\ \bibinfo {pages} {435--439}
  (\bibinfo {year} {2017})}\BibitemShut {NoStop}%
\bibitem [{\citenamefont {Rodriguez-Nieva}\ and\ \citenamefont
  {Scheurer}(2019)}]{rodriguez2019identifying}%
  \BibitemOpen
  \bibfield  {author} {\bibinfo {author} {\bibfnamefont {Joaquin~F}\
  \bibnamefont {Rodriguez-Nieva}}\ and\ \bibinfo {author} {\bibfnamefont
  {Mathias~S}\ \bibnamefont {Scheurer}},\ }\bibfield  {title} {\enquote
  {\bibinfo {title} {Identifying topological order through unsupervised machine
  learning},}\ }\href {10.1038/s41567-019-0545-1} {\bibfield  {journal}
  {\bibinfo  {journal} {Nature Physics}\ }\textbf {\bibinfo {volume} {15}},\
  \bibinfo {pages} {790--795} (\bibinfo {year} {2019})}\BibitemShut {NoStop}%
\bibitem [{\citenamefont {Scheurer}\ and\ \citenamefont
  {Slager}(2020)}]{PhysRevLett.124.226401}%
  \BibitemOpen
  \bibfield  {author} {\bibinfo {author} {\bibfnamefont {Mathias~S.}\
  \bibnamefont {Scheurer}}\ and\ \bibinfo {author} {\bibfnamefont {Robert-Jan}\
  \bibnamefont {Slager}},\ }\bibfield  {title} {\enquote {\bibinfo {title}
  {Unsupervised machine learning and band topology},}\ }\href {\doibase
  10.1103/PhysRevLett.124.226401} {\bibfield  {journal} {\bibinfo  {journal}
  {Phys. Rev. Lett.}\ }\textbf {\bibinfo {volume} {124}},\ \bibinfo {pages}
  {226401} (\bibinfo {year} {2020})}\BibitemShut {NoStop}%
\bibitem [{\citenamefont {Long}\ \emph {et~al.}(2020)\citenamefont {Long},
  \citenamefont {Ren},\ and\ \citenamefont {Chen}}]{PhysRevLett.124.185501}%
  \BibitemOpen
  \bibfield  {author} {\bibinfo {author} {\bibfnamefont {Yang}\ \bibnamefont
  {Long}}, \bibinfo {author} {\bibfnamefont {Jie}\ \bibnamefont {Ren}}, \ and\
  \bibinfo {author} {\bibfnamefont {Hong}\ \bibnamefont {Chen}},\ }\bibfield
  {title} {\enquote {\bibinfo {title} {Unsupervised manifold clustering of
  topological phononics},}\ }\href {\doibase 10.1103/PhysRevLett.124.185501}
  {\bibfield  {journal} {\bibinfo  {journal} {Phys. Rev. Lett.}\ }\textbf
  {\bibinfo {volume} {124}},\ \bibinfo {pages} {185501} (\bibinfo {year}
  {2020})}\BibitemShut {NoStop}%
\bibitem [{\citenamefont {Che}\ \emph {et~al.}(2020)\citenamefont {Che},
  \citenamefont {Gneiting}, \citenamefont {Liu},\ and\ \citenamefont
  {Nori}}]{che2020topological}%
  \BibitemOpen
  \bibfield  {author} {\bibinfo {author} {\bibfnamefont {Yanming}\ \bibnamefont
  {Che}}, \bibinfo {author} {\bibfnamefont {Clemens}\ \bibnamefont {Gneiting}},
  \bibinfo {author} {\bibfnamefont {Tao}\ \bibnamefont {Liu}}, \ and\ \bibinfo
  {author} {\bibfnamefont {Franco}\ \bibnamefont {Nori}},\ }\bibfield  {title}
  {\enquote {\bibinfo {title} {Topological quantum phase transitions retrieved
  from manifold learning},}\ }\href {https://arxiv.org/abs/2002.02363}
  {\bibfield  {journal} {\bibinfo  {journal} {arXiv:2002.02363}\ } (\bibinfo
  {year} {2020})}\BibitemShut {NoStop}%
\bibitem [{\citenamefont {Kharkov}\ \emph {et~al.}(2020)\citenamefont
  {Kharkov}, \citenamefont {Sotskov}, \citenamefont {Karazeev}, \citenamefont
  {Kiktenko},\ and\ \citenamefont {Fedorov}}]{PhysRevB.101.064406}%
  \BibitemOpen
  \bibfield  {author} {\bibinfo {author} {\bibfnamefont {Y.~A.}\ \bibnamefont
  {Kharkov}}, \bibinfo {author} {\bibfnamefont {V.~E.}\ \bibnamefont
  {Sotskov}}, \bibinfo {author} {\bibfnamefont {A.~A.}\ \bibnamefont
  {Karazeev}}, \bibinfo {author} {\bibfnamefont {E.~O.}\ \bibnamefont
  {Kiktenko}}, \ and\ \bibinfo {author} {\bibfnamefont {A.~K.}\ \bibnamefont
  {Fedorov}},\ }\bibfield  {title} {\enquote {\bibinfo {title} {Revealing
  quantum chaos with machine learning},}\ }\href {\doibase
  10.1103/PhysRevB.101.064406} {\bibfield  {journal} {\bibinfo  {journal}
  {Phys. Rev. B}\ }\textbf {\bibinfo {volume} {101}},\ \bibinfo {pages}
  {064406} (\bibinfo {year} {2020})}\BibitemShut {NoStop}%
\bibitem [{\citenamefont {Torlai}\ \emph {et~al.}(2018)\citenamefont {Torlai},
  \citenamefont {Mazzola}, \citenamefont {Carrasquilla}, \citenamefont
  {Troyer}, \citenamefont {Melko},\ and\ \citenamefont {Carleo}}]{Torlai2018}%
  \BibitemOpen
  \bibfield  {author} {\bibinfo {author} {\bibfnamefont {Giacomo}\ \bibnamefont
  {Torlai}}, \bibinfo {author} {\bibfnamefont {Guglielmo}\ \bibnamefont
  {Mazzola}}, \bibinfo {author} {\bibfnamefont {Juan}\ \bibnamefont
  {Carrasquilla}}, \bibinfo {author} {\bibfnamefont {Matthias}\ \bibnamefont
  {Troyer}}, \bibinfo {author} {\bibfnamefont {Roger}\ \bibnamefont {Melko}}, \
  and\ \bibinfo {author} {\bibfnamefont {Giuseppe}\ \bibnamefont {Carleo}},\
  }\bibfield  {title} {\enquote {\bibinfo {title} {Neural-network quantum state
  tomography},}\ }\href {\doibase 10.1038/s41567-018-0048-5} {\bibfield
  {journal} {\bibinfo  {journal} {Nature Physics}\ }\textbf {\bibinfo {volume}
  {14}},\ \bibinfo {pages} {447--450} (\bibinfo {year} {2018})}\BibitemShut
  {NoStop}%
\bibitem [{\citenamefont {Zhang}\ and\ \citenamefont
  {Kim}(2017)}]{PhysRevLett.118.216401}%
  \BibitemOpen
  \bibfield  {author} {\bibinfo {author} {\bibfnamefont {Yi}~\bibnamefont
  {Zhang}}\ and\ \bibinfo {author} {\bibfnamefont {Eun-Ah}\ \bibnamefont
  {Kim}},\ }\bibfield  {title} {\enquote {\bibinfo {title} {Quantum loop
  topography for machine learning},}\ }\href {\doibase
  10.1103/PhysRevLett.118.216401} {\bibfield  {journal} {\bibinfo  {journal}
  {Phys. Rev. Lett.}\ }\textbf {\bibinfo {volume} {118}},\ \bibinfo {pages}
  {216401} (\bibinfo {year} {2017})}\BibitemShut {NoStop}%
\bibitem [{\citenamefont {Torlai}\ \emph {et~al.}(2019)\citenamefont {Torlai},
  \citenamefont {Timar}, \citenamefont {van Nieuwenburg}, \citenamefont
  {Levine}, \citenamefont {Omran}, \citenamefont {Keesling}, \citenamefont
  {Bernien}, \citenamefont {Greiner}, \citenamefont
  {Vuleti\ifmmode~\acute{c}\else \'{c}\fi{}}, \citenamefont {Lukin},
  \citenamefont {Melko},\ and\ \citenamefont
  {Endres}}]{PhysRevLett.123.230504}%
  \BibitemOpen
  \bibfield  {author} {\bibinfo {author} {\bibfnamefont {Giacomo}\ \bibnamefont
  {Torlai}}, \bibinfo {author} {\bibfnamefont {Brian}\ \bibnamefont {Timar}},
  \bibinfo {author} {\bibfnamefont {Evert P.~L.}\ \bibnamefont {van
  Nieuwenburg}}, \bibinfo {author} {\bibfnamefont {Harry}\ \bibnamefont
  {Levine}}, \bibinfo {author} {\bibfnamefont {Ahmed}\ \bibnamefont {Omran}},
  \bibinfo {author} {\bibfnamefont {Alexander}\ \bibnamefont {Keesling}},
  \bibinfo {author} {\bibfnamefont {Hannes}\ \bibnamefont {Bernien}}, \bibinfo
  {author} {\bibfnamefont {Markus}\ \bibnamefont {Greiner}}, \bibinfo {author}
  {\bibfnamefont {Vladan}\ \bibnamefont {Vuleti\ifmmode~\acute{c}\else
  \'{c}\fi{}}}, \bibinfo {author} {\bibfnamefont {Mikhail~D.}\ \bibnamefont
  {Lukin}}, \bibinfo {author} {\bibfnamefont {Roger~G.}\ \bibnamefont {Melko}},
  \ and\ \bibinfo {author} {\bibfnamefont {Manuel}\ \bibnamefont {Endres}},\
  }\bibfield  {title} {\enquote {\bibinfo {title} {Integrating neural networks
  with a quantum simulator for state reconstruction},}\ }\href {\doibase
  10.1103/PhysRevLett.123.230504} {\bibfield  {journal} {\bibinfo  {journal}
  {Phys. Rev. Lett.}\ }\textbf {\bibinfo {volume} {123}},\ \bibinfo {pages}
  {230504} (\bibinfo {year} {2019})}\BibitemShut {NoStop}%
\bibitem [{\citenamefont {Gong}\ \emph {et~al.}(2018)\citenamefont {Gong},
  \citenamefont {Ashida}, \citenamefont {Kawabata}, \citenamefont {Takasan},
  \citenamefont {Higashikawa},\ and\ \citenamefont {Ueda}}]{PhysRevX.8.031079}%
  \BibitemOpen
  \bibfield  {author} {\bibinfo {author} {\bibfnamefont {Zongping}\
  \bibnamefont {Gong}}, \bibinfo {author} {\bibfnamefont {Yuto}\ \bibnamefont
  {Ashida}}, \bibinfo {author} {\bibfnamefont {Kohei}\ \bibnamefont
  {Kawabata}}, \bibinfo {author} {\bibfnamefont {Kazuaki}\ \bibnamefont
  {Takasan}}, \bibinfo {author} {\bibfnamefont {Sho}\ \bibnamefont
  {Higashikawa}}, \ and\ \bibinfo {author} {\bibfnamefont {Masahito}\
  \bibnamefont {Ueda}},\ }\bibfield  {title} {\enquote {\bibinfo {title}
  {Topological phases of non-hermitian systems},}\ }\href {\doibase
  10.1103/PhysRevX.8.031079} {\bibfield  {journal} {\bibinfo  {journal} {Phys.
  Rev. X}\ }\textbf {\bibinfo {volume} {8}},\ \bibinfo {pages} {031079}
  (\bibinfo {year} {2018})}\BibitemShut {NoStop}%
\bibitem [{\citenamefont {Bukov}\ \emph {et~al.}(2018)\citenamefont {Bukov},
  \citenamefont {Day}, \citenamefont {Sels}, \citenamefont {Weinberg},
  \citenamefont {Polkovnikov},\ and\ \citenamefont
  {Mehta}}]{PhysRevX.8.031086}%
  \BibitemOpen
  \bibfield  {author} {\bibinfo {author} {\bibfnamefont {Marin}\ \bibnamefont
  {Bukov}}, \bibinfo {author} {\bibfnamefont {Alexandre G.~R.}\ \bibnamefont
  {Day}}, \bibinfo {author} {\bibfnamefont {Dries}\ \bibnamefont {Sels}},
  \bibinfo {author} {\bibfnamefont {Phillip}\ \bibnamefont {Weinberg}},
  \bibinfo {author} {\bibfnamefont {Anatoli}\ \bibnamefont {Polkovnikov}}, \
  and\ \bibinfo {author} {\bibfnamefont {Pankaj}\ \bibnamefont {Mehta}},\
  }\bibfield  {title} {\enquote {\bibinfo {title} {Reinforcement learning in
  different phases of quantum control},}\ }\href {\doibase
  10.1103/PhysRevX.8.031086} {\bibfield  {journal} {\bibinfo  {journal} {Phys.
  Rev. X}\ }\textbf {\bibinfo {volume} {8}},\ \bibinfo {pages} {031086}
  (\bibinfo {year} {2018})}\BibitemShut {NoStop}%
\bibitem [{\citenamefont {Dalgaard}\ \emph {et~al.}(2020)\citenamefont
  {Dalgaard}, \citenamefont {Motzoi}, \citenamefont {Sorensen},\ and\
  \citenamefont {Sherson}}]{dalgaard2020global}%
  \BibitemOpen
  \bibfield  {author} {\bibinfo {author} {\bibfnamefont {Mogens}\ \bibnamefont
  {Dalgaard}}, \bibinfo {author} {\bibfnamefont {Felix}\ \bibnamefont
  {Motzoi}}, \bibinfo {author} {\bibfnamefont {Jens~Jakob}\ \bibnamefont
  {Sorensen}}, \ and\ \bibinfo {author} {\bibfnamefont {Jacob}\ \bibnamefont
  {Sherson}},\ }\bibfield  {title} {\enquote {\bibinfo {title} {Global
  optimization of quantum dynamics with alphazero deep exploration},}\ }\href
  {\doibase 10.1038/s41534-019-0241-0} {\bibfield  {journal} {\bibinfo
  {journal} {npj Quantum Information}\ }\textbf {\bibinfo {volume} {6}}
  (\bibinfo {year} {2020}),\ 10.1038/s41534-019-0241-0}\BibitemShut {NoStop}%
\bibitem [{\citenamefont {Torlai}\ and\ \citenamefont
  {Melko}(2017)}]{PhysRevLett.119.030501}%
  \BibitemOpen
  \bibfield  {author} {\bibinfo {author} {\bibfnamefont {Giacomo}\ \bibnamefont
  {Torlai}}\ and\ \bibinfo {author} {\bibfnamefont {Roger~G.}\ \bibnamefont
  {Melko}},\ }\bibfield  {title} {\enquote {\bibinfo {title} {Neural decoder
  for topological codes},}\ }\href {\doibase 10.1103/PhysRevLett.119.030501}
  {\bibfield  {journal} {\bibinfo  {journal} {Phys. Rev. Lett.}\ }\textbf
  {\bibinfo {volume} {119}},\ \bibinfo {pages} {030501} (\bibinfo {year}
  {2017})}\BibitemShut {NoStop}%
\bibitem [{\citenamefont {Baireuther}\ \emph {et~al.}(2018)\citenamefont
  {Baireuther}, \citenamefont {O'Brien}, \citenamefont {Tarasinski},\ and\
  \citenamefont {Beenakker}}]{Baireuther2018machinelearning}%
  \BibitemOpen
  \bibfield  {author} {\bibinfo {author} {\bibfnamefont {Paul}\ \bibnamefont
  {Baireuther}}, \bibinfo {author} {\bibfnamefont {Thomas~E.}\ \bibnamefont
  {O'Brien}}, \bibinfo {author} {\bibfnamefont {Brian}\ \bibnamefont
  {Tarasinski}}, \ and\ \bibinfo {author} {\bibfnamefont {Carlo W.~J.}\
  \bibnamefont {Beenakker}},\ }\bibfield  {title} {\enquote {\bibinfo {title}
  {Machine-learning-assisted correction of correlated qubit errors in a
  topological code},}\ }\href {\doibase 10.22331/q-2018-01-29-48} {\bibfield
  {journal} {\bibinfo  {journal} {{Quantum}}\ }\textbf {\bibinfo {volume}
  {2}},\ \bibinfo {pages} {48} (\bibinfo {year} {2018})}\BibitemShut {NoStop}%
\bibitem [{\citenamefont {{Sweke}}\ \emph {et~al.}(2018)\citenamefont
  {{Sweke}}, \citenamefont {{Kesselring}}, \citenamefont {{van Nieuwenburg}},\
  and\ \citenamefont {{Eisert}}}]{2018arXiv181007207S}%
  \BibitemOpen
  \bibfield  {author} {\bibinfo {author} {\bibfnamefont {Ryan}\ \bibnamefont
  {{Sweke}}}, \bibinfo {author} {\bibfnamefont {Markus~S.}\ \bibnamefont
  {{Kesselring}}}, \bibinfo {author} {\bibfnamefont {Evert P.~L.}\ \bibnamefont
  {{van Nieuwenburg}}}, \ and\ \bibinfo {author} {\bibfnamefont {Jens}\
  \bibnamefont {{Eisert}}},\ }\bibfield  {title} {\enquote {\bibinfo {title}
  {{Reinforcement Learning Decoders for Fault-Tolerant Quantum Computation}},}\
  }\href {https://ui.adsabs.harvard.edu/abs/2018arXiv181007207S} {\bibfield
  {journal} {\bibinfo  {journal} {arXiv:1810.07207}\ } (\bibinfo {year}
  {2018})}\BibitemShut {NoStop}%
\bibitem [{\citenamefont {Andreasson}\ \emph {et~al.}(2019)\citenamefont
  {Andreasson}, \citenamefont {Johansson}, \citenamefont {Liljestrand},\ and\
  \citenamefont {Granath}}]{Andreasson2019quantumerror}%
  \BibitemOpen
  \bibfield  {author} {\bibinfo {author} {\bibfnamefont {Philip}\ \bibnamefont
  {Andreasson}}, \bibinfo {author} {\bibfnamefont {Joel}\ \bibnamefont
  {Johansson}}, \bibinfo {author} {\bibfnamefont {Simon}\ \bibnamefont
  {Liljestrand}}, \ and\ \bibinfo {author} {\bibfnamefont {Mats}\ \bibnamefont
  {Granath}},\ }\bibfield  {title} {\enquote {\bibinfo {title} {Quantum error
  correction for the toric code using deep reinforcement learning},}\ }\href
  {\doibase 10.22331/q-2019-09-02-183} {\bibfield  {journal} {\bibinfo
  {journal} {{Quantum}}\ }\textbf {\bibinfo {volume} {3}},\ \bibinfo {pages}
  {183} (\bibinfo {year} {2019})}\BibitemShut {NoStop}%
\bibitem [{\citenamefont {Nautrup}\ \emph {et~al.}(2019)\citenamefont
  {Nautrup}, \citenamefont {Delfosse}, \citenamefont {Dunjko}, \citenamefont
  {Briegel},\ and\ \citenamefont {Friis}}]{nautrup2019optimizing}%
  \BibitemOpen
  \bibfield  {author} {\bibinfo {author} {\bibfnamefont {Hendrik~Poulsen}\
  \bibnamefont {Nautrup}}, \bibinfo {author} {\bibfnamefont {Nicolas}\
  \bibnamefont {Delfosse}}, \bibinfo {author} {\bibfnamefont {Vedran}\
  \bibnamefont {Dunjko}}, \bibinfo {author} {\bibfnamefont {Hans~J}\
  \bibnamefont {Briegel}}, \ and\ \bibinfo {author} {\bibfnamefont {Nicolai}\
  \bibnamefont {Friis}},\ }\bibfield  {title} {\enquote {\bibinfo {title}
  {Optimizing quantum error correction codes with reinforcement learning},}\
  }\href {\doibase 0.22331/q-2019-12-16-215} {\bibfield  {journal} {\bibinfo
  {journal} {Quantum}\ }\textbf {\bibinfo {volume} {3}},\ \bibinfo {pages}
  {215} (\bibinfo {year} {2019})}\BibitemShut {NoStop}%
\bibitem [{\citenamefont {Valenti}\ \emph {et~al.}(2019)\citenamefont
  {Valenti}, \citenamefont {van Nieuwenburg}, \citenamefont {Huber},\ and\
  \citenamefont {Greplova}}]{PhysRevResearch.1.033092}%
  \BibitemOpen
  \bibfield  {author} {\bibinfo {author} {\bibfnamefont {Agnes}\ \bibnamefont
  {Valenti}}, \bibinfo {author} {\bibfnamefont {Evert}\ \bibnamefont {van
  Nieuwenburg}}, \bibinfo {author} {\bibfnamefont {Sebastian}\ \bibnamefont
  {Huber}}, \ and\ \bibinfo {author} {\bibfnamefont {Eliska}\ \bibnamefont
  {Greplova}},\ }\bibfield  {title} {\enquote {\bibinfo {title} {Hamiltonian
  learning for quantum error correction},}\ }\href {\doibase
  10.1103/PhysRevResearch.1.033092} {\bibfield  {journal} {\bibinfo  {journal}
  {Phys. Rev. Research}\ }\textbf {\bibinfo {volume} {1}},\ \bibinfo {pages}
  {033092} (\bibinfo {year} {2019})}\BibitemShut {NoStop}%
\bibitem [{\citenamefont {Hughes}\ \emph {et~al.}(2011)\citenamefont {Hughes},
  \citenamefont {Prodan},\ and\ \citenamefont {Bernevig}}]{Hughes_Inversion}%
  \BibitemOpen
  \bibfield  {author} {\bibinfo {author} {\bibfnamefont {Taylor~L.}\
  \bibnamefont {Hughes}}, \bibinfo {author} {\bibfnamefont {Emil}\ \bibnamefont
  {Prodan}}, \ and\ \bibinfo {author} {\bibfnamefont {B.~Andrei}\ \bibnamefont
  {Bernevig}},\ }\bibfield  {title} {\enquote {\bibinfo {title}
  {Inversion-symmetric topological insulators},}\ }\href {\doibase
  10.1103/PhysRevB.83.245132} {\bibfield  {journal} {\bibinfo  {journal} {Phys.
  Rev. B}\ }\textbf {\bibinfo {volume} {83}},\ \bibinfo {pages} {245132}
  (\bibinfo {year} {2011})}\BibitemShut {NoStop}%
\bibitem [{\citenamefont {Freed}\ and\ \citenamefont
  {Moore}(2013)}]{Freed2013}%
  \BibitemOpen
  \bibfield  {author} {\bibinfo {author} {\bibfnamefont {Daniel~S.}\
  \bibnamefont {Freed}}\ and\ \bibinfo {author} {\bibfnamefont {Gregory~W.}\
  \bibnamefont {Moore}},\ }\bibfield  {title} {\enquote {\bibinfo {title}
  {Twisted equivariant matter},}\ }\href {\doibase 10.1007/s00023-013-0236-x}
  {\bibfield  {journal} {\bibinfo  {journal} {Annales Henri Poincar\'e}\
  }\textbf {\bibinfo {volume} {14}},\ \bibinfo {pages} {1927--2023} (\bibinfo
  {year} {2013})}\BibitemShut {NoStop}%
\bibitem [{\citenamefont {Kitaev}(2009)}]{doi:10.1063/1.3149495}%
  \BibitemOpen
  \bibfield  {author} {\bibinfo {author} {\bibfnamefont {Alexei}\ \bibnamefont
  {Kitaev}},\ }\bibfield  {title} {\enquote {\bibinfo {title} {Periodic table
  for topological insulators and superconductors},}\ }\href {\doibase
  10.1063/1.3149495} {\bibfield  {journal} {\bibinfo  {journal} {AIP Conference
  Proceedings}\ }\textbf {\bibinfo {volume} {1134}},\ \bibinfo {pages} {22--30}
  (\bibinfo {year} {2009})}\BibitemShut {NoStop}%
\bibitem [{\citenamefont {{Lu}}\ and\ \citenamefont
  {{Lee}}(2014)}]{2014arXiv1403.5558L}%
  \BibitemOpen
  \bibfield  {author} {\bibinfo {author} {\bibfnamefont {Yuan-Ming}\
  \bibnamefont {{Lu}}}\ and\ \bibinfo {author} {\bibfnamefont {Dung-Hai}\
  \bibnamefont {{Lee}}},\ }\bibfield  {title} {\enquote {\bibinfo {title}
  {{Inversion symmetry protected topological insulators and
  superconductors}},}\ }\href
  {https://ui.adsabs.harvard.edu/abs/2014arXiv1403.5558L} {\bibfield  {journal}
  {\bibinfo  {journal} {arXiv:1403.5558}\ } (\bibinfo {year}
  {2014})}\BibitemShut {NoStop}%
\bibitem [{\citenamefont {Lutchyn}\ \emph {et~al.}(2018)\citenamefont
  {Lutchyn}, \citenamefont {Bakkers}, \citenamefont {Kouwenhoven},
  \citenamefont {Krogstrup}, \citenamefont {Marcus},\ and\ \citenamefont
  {Oreg}}]{Lutchyn2018}%
  \BibitemOpen
  \bibfield  {author} {\bibinfo {author} {\bibfnamefont {R.~M.}\ \bibnamefont
  {Lutchyn}}, \bibinfo {author} {\bibfnamefont {E.~P. A.~M.}\ \bibnamefont
  {Bakkers}}, \bibinfo {author} {\bibfnamefont {L.~P.}\ \bibnamefont
  {Kouwenhoven}}, \bibinfo {author} {\bibfnamefont {P.}~\bibnamefont
  {Krogstrup}}, \bibinfo {author} {\bibfnamefont {C.~M.}\ \bibnamefont
  {Marcus}}, \ and\ \bibinfo {author} {\bibfnamefont {Y.}~\bibnamefont
  {Oreg}},\ }\bibfield  {title} {\enquote {\bibinfo {title} {Majorana zero
  modes in superconductor - semiconductor heterostructures},}\ }\href {\doibase
  10.1038/s41578-018-0003-1} {\bibfield  {journal} {\bibinfo  {journal} {Nature
  Reviews Materials}\ }\textbf {\bibinfo {volume} {3}},\ \bibinfo {pages}
  {52--68} (\bibinfo {year} {2018})}\BibitemShut {NoStop}%
\bibitem [{\citenamefont {Budich}\ and\ \citenamefont
  {Ardonne}(2013)}]{PhysRevB.88.075419}%
  \BibitemOpen
  \bibfield  {author} {\bibinfo {author} {\bibfnamefont {Jan~Carl}\
  \bibnamefont {Budich}}\ and\ \bibinfo {author} {\bibfnamefont {Eddy}\
  \bibnamefont {Ardonne}},\ }\bibfield  {title} {\enquote {\bibinfo {title}
  {Equivalent topological invariants for one-dimensional majorana wires in
  symmetry class $d$},}\ }\href {\doibase 10.1103/PhysRevB.88.075419}
  {\bibfield  {journal} {\bibinfo  {journal} {Phys. Rev. B}\ }\textbf {\bibinfo
  {volume} {88}},\ \bibinfo {pages} {075419} (\bibinfo {year}
  {2013})}\BibitemShut {NoStop}%
\bibitem [{\citenamefont {Kitagawa}\ \emph {et~al.}(2010)\citenamefont
  {Kitagawa}, \citenamefont {Berg}, \citenamefont {Rudner},\ and\ \citenamefont
  {Demler}}]{PhysRevB.82.235114}%
  \BibitemOpen
  \bibfield  {author} {\bibinfo {author} {\bibfnamefont {Takuya}\ \bibnamefont
  {Kitagawa}}, \bibinfo {author} {\bibfnamefont {Erez}\ \bibnamefont {Berg}},
  \bibinfo {author} {\bibfnamefont {Mark}\ \bibnamefont {Rudner}}, \ and\
  \bibinfo {author} {\bibfnamefont {Eugene}\ \bibnamefont {Demler}},\
  }\bibfield  {title} {\enquote {\bibinfo {title} {Topological characterization
  of periodically driven quantum systems},}\ }\href {\doibase
  10.1103/PhysRevB.82.235114} {\bibfield  {journal} {\bibinfo  {journal} {Phys.
  Rev. B}\ }\textbf {\bibinfo {volume} {82}},\ \bibinfo {pages} {235114}
  (\bibinfo {year} {2010})}\BibitemShut {NoStop}%
\bibitem [{\citenamefont {Lindner}\ \emph {et~al.}(2011)\citenamefont
  {Lindner}, \citenamefont {Refael},\ and\ \citenamefont
  {Galitski}}]{Lindner2011}%
  \BibitemOpen
  \bibfield  {author} {\bibinfo {author} {\bibfnamefont {Netanel~H.}\
  \bibnamefont {Lindner}}, \bibinfo {author} {\bibfnamefont {Gil}\ \bibnamefont
  {Refael}}, \ and\ \bibinfo {author} {\bibfnamefont {Victor}\ \bibnamefont
  {Galitski}},\ }\bibfield  {title} {\enquote {\bibinfo {title} {Floquet
  topological insulator in semiconductor quantum wells},}\ }\href {\doibase
  10.1038/nphys1926} {\bibfield  {journal} {\bibinfo  {journal} {Nature
  Physics}\ }\textbf {\bibinfo {volume} {7}},\ \bibinfo {pages} {490--495}
  (\bibinfo {year} {2011})}\BibitemShut {NoStop}%
\bibitem [{\citenamefont {Kawabata}\ \emph {et~al.}(2019)\citenamefont
  {Kawabata}, \citenamefont {Shiozaki}, \citenamefont {Ueda},\ and\
  \citenamefont {Sato}}]{PhysRevX.9.041015}%
  \BibitemOpen
  \bibfield  {author} {\bibinfo {author} {\bibfnamefont {Kohei}\ \bibnamefont
  {Kawabata}}, \bibinfo {author} {\bibfnamefont {Ken}\ \bibnamefont
  {Shiozaki}}, \bibinfo {author} {\bibfnamefont {Masahito}\ \bibnamefont
  {Ueda}}, \ and\ \bibinfo {author} {\bibfnamefont {Masatoshi}\ \bibnamefont
  {Sato}},\ }\bibfield  {title} {\enquote {\bibinfo {title} {Symmetry and
  topology in non-hermitian physics},}\ }\href {\doibase
  10.1103/PhysRevX.9.041015} {\bibfield  {journal} {\bibinfo  {journal} {Phys.
  Rev. X}\ }\textbf {\bibinfo {volume} {9}},\ \bibinfo {pages} {041015}
  (\bibinfo {year} {2019})}\BibitemShut {NoStop}%
\bibitem [{\citenamefont {Fidkowski}\ and\ \citenamefont
  {Kitaev}(2011)}]{PhysRevB.83.075103}%
  \BibitemOpen
  \bibfield  {author} {\bibinfo {author} {\bibfnamefont {Lukasz}\ \bibnamefont
  {Fidkowski}}\ and\ \bibinfo {author} {\bibfnamefont {Alexei}\ \bibnamefont
  {Kitaev}},\ }\bibfield  {title} {\enquote {\bibinfo {title} {Topological
  phases of fermions in one dimension},}\ }\href {\doibase
  10.1103/PhysRevB.83.075103} {\bibfield  {journal} {\bibinfo  {journal} {Phys.
  Rev. B}\ }\textbf {\bibinfo {volume} {83}},\ \bibinfo {pages} {075103}
  (\bibinfo {year} {2011})}\BibitemShut {NoStop}%
\end{thebibliography}%

\end{document}